\documentclass[preprint,aps,showpacs,nofootinbib]{revtex4}

\usepackage{graphicx,epsfig,amssymb,amsmath,color}

\def\bea{\begin{eqnarray}}
\def\eea{\end{eqnarray}}
\def\beq{\begin{equation}}
\def\eeq{\end{equation}}

\begin{document}
\draft
\tighten
\preprint{CTPU-14-10, DESY-14-103, TU-974}
\title{\Large \bf
Higgs Mixing in the NMSSM and Light Higgsinos
}
\author{
Kwang Sik Jeong$^{\,a,b}$\footnote{e-mail: ksjeong@ibs.re.kr},
Yutaro Shoji$^{\,c}$\footnote{e-mail: yshoji@tuhep.phys.tohoku.ac.jp},
Masahiro Yamaguchi$^{\,c}$\footnote{e-mail: yama@tuhep.phys.tohoku.ac.jp}
}
\affiliation{
$^a$ DESY, Notkestrasse 85, 22607 Hamburg, Germany \\
$^b$ Center for Theoretical Physics of the Universe, IBS, Daejeon 305-811, Korea \\
$^c$ Department of Physics, Tohoku University, Sendai 980-8578, Japan
}

\vspace{2cm}

\begin{abstract}

We explore the effects of Higgs mixing in the general next-to-minimal supersymmetric
Standard Model (NMSSM).
Extended to include a gauge singlet, the Higgs sector can naturally explain the observed
Higgs boson mass in TeV scale supersymmetry without invoking large stop mixing.
This is particularly the case when the singlet scalar is light so that singlet-doublet
mixing increases the mass of the SM-like Higgs boson.
In such a case the Higgs mixing has interesting implications following from the fact that
the higgsino mass parameter and the singlet coupling to Higgs bilinear crucially depend
on the Higgs boson masses and mixing angles.
For the mixing compatible with the current LHC data on the Higgs signal rates,
the higgsinos are required to be relatively light, around or below a few hundred GeV,
as long as the heavy doublet Higgs boson has a mass smaller than about
$250\sqrt{\tan\beta}$~GeV and the singlet-like Higgs boson is consistent with the LEP
constraint.
In addition, the Higgs coupling to photons can receive a sizable contribution of
either sign from the charged-higgsino loops combined with singlet-doublet mixing.

\end{abstract}

\maketitle

\section{Introduction}

Supersymmetry (SUSY) provides a natural solution to the gauge hierarchy problem of
the Standard Model (SM), and is a leading candidate for new physics at
the TeV scale \cite{susy-sm}.
To be consistent with the LHC data, the supersymmetric extension of the SM should
accommodate a scalar boson that has a mass near $126$~GeV and behaves like
the SM Higgs boson \cite{ATLAS-Higgs,CMS-Higgs}.
The minimal supersymmetric SM (MSSM) can explain the observed mass if one considers
heavy stops above 7~TeV or large stop mixing, which however would cause
severe fine-tuning in the electroweak symmetry breaking.
Such a difficulty can be avoided in the next-to-minimal supersymmetric SM (NMSSM)
\cite{review-NMSSM,fine-tuning}, where the Higgs sector is extended to include a gauge singlet
$S$ interacting with the Higgs doublets via the superpotential coupling,
$\lambda SH_uH_d$.
The NMSSM can provide a larger mass to the SM-like Higgs boson because there
are additional tree-level contributions from the $F$-term scalar potential
$\lambda^2 |H_uH_d|^2$, and from the singlet-doublet mixing.

The Higgs sector has a rich structure in the NMSSM due to the singlet scalar.
In particular, singlet-doublet scalar mixing can have interesting phenomenological
consequences \cite{diphoton-NMSSM,mixing-mh,diphoton-enhancement-nmssm,nmssm-higgs,nmssm-pheno}.
It increases the mass of the SM-like Higgs boson if the singlet scalar is light,
which does not require low $\tan\beta$ or sizable $\lambda$ differently from
the singlet $F$-term contribution.
In addition, the singlet-doublet mixing induces a Higgs coupling to photons of either
sign through the charged-higgsino loops \cite{diphoton-NMSSM}.
In this paper we study the effects of Higgs mixing in the NMSSM where
the lightest CP-even neutral scalar is singlet-like, i.e.~lighter than the SM-like
Higgs boson $h$, under the assumption that all the superparticles have masses around
or below TeV as would be required to solve the hierarchy problem.
Since our results apply to any NMSSM model, we will not specify the exact form of
the singlet superpotential or the mediation mechanism of SUSY breaking.

In the presence of scalar mixing, $h$ has properties deviated from those of the SM
Higgs boson.
However sizable mixing is still compatible with the current experimental data on
the Higgs signal rates in TeV scale SUSY, especially when the singlet-doublet
mixing increases the mass of $h$.
The scalar mixing depends on the coupling $\lambda$ and the higgsino mass parameter
$\mu$.
This implies that the LHC and LEP constraints on scalar mixing are converted into
the constraints on $\lambda$ and $\mu$, and vice versa.
Interestingly it turns out that higgsinos are required to be light as $h$ becomes more
SM-like in NMSSM models where the singlet-like Higgs boson is lighter than $h$.
We find that higgsinos have masses around or below a few hundred GeV for the scalar
mixing compatible with the LHC results on Higgs signal rates, as long as the heavy
doublet Higgs boson has a mass less than about $250\sqrt{\tan\beta}$~GeV.
The upper bound on $\mu$ also indicates that the heavy doublet Higgs boson should
have a large mass at large $\tan\beta$ in order to allow $|\mu|>100$~GeV in the viable
region of scalar mixing, as suggested by the LEP bound on the chargino mass \cite{PDG}.

This paper is organized as follows.
In section \ref{sec:Higgs-NMSSM} we discuss some generic features of scalar mixing
in the NMSSM Higgs sector, and the mixing effects on the SM-like Higgs boson.
In section \ref{sec:NMSSM-light-singlet}, we consider the case with a light singlet
scalar, for which the SM-like Higgs boson can obtain the required mass via
singlet-doublet mixing in TeV scale SUSY without large stop mixing.
We will examine the range of mixing angles compatible with the current LHC and LEP data,
and then discuss the implications of scalar mixing on the higgsino properties
by using the fact that the Higgs boson masses and scalar mixing angles crucially depend
on $\lambda$ and $\mu$.
Section \ref{sec:conclusion} is the conclusions.

\section{Higgs properties in the NMSSM}\label{sec:Higgs-NMSSM}

In this section we briefly discuss how the Higgs sector is modified by
the gauge singlet in the NMSSM.
Then we summarize experimental and theoretical constraints on the Higgs sector,
paying attention to how to arrange a 126~GeV Higgs boson with
SM-like properties in TeV scale SUSY.

\subsection{Higgs sector}

A singlet extension of the MSSM can always be described by the superpotential
\bea
\label{nmssm-superpotential}
W = \lambda S H_u H_d + f(S) + (\mbox{MSSM Yukawa terms}),
\eea
in the field basis where the gauge singlet $S$ has a minimal K\"ahler potential,
$|S|^2$.
Here the singlet superpotential $f$ is needed to avoid a phenomenologically unacceptable
visible axion, and it has no dependence on the MSSM superfields at the renormalizable level.
There are various NMSSM models classified by the form of $f$.
In this paper we do not specify the form of $f$ as our results do not depend
much on it, but we will assume no CP violation in the Higgs sector.

After electroweak symmetry breaking (EWSB), the doublet Higgs bosons mix with the singlet
boson via the couplings
\bea
-{\cal L}_{\rm mix} =
\lambda^2 |S|^2 (|H_u|^2 + |H_d|^2)
+ \Big( A_\lambda \lambda S H_u H_d +  (\partial_S f)^* \lambda H_u H_d + {\rm h.c.}
\Big),
\eea
where $A_\lambda$ is the soft SUSY breaking trilinear parameter.
Using the EWSB conditions, one can find that the mass squared matrix for the neutral
CP-even scalar bosons is written
\bea
\label{mass-squared}
\begin{pmatrix}
 m_0^2+(\lambda^2v^2-m_Z^2)\sin^22\beta&-(\lambda^2v^2-m_Z^2)\sin2\beta\cos2\beta&\lambda
 v(2\mu-\Lambda\sin2\beta)\\
 -(\lambda^2v^2-m_Z^2)\sin2\beta\cos2\beta&-(\lambda^2v^2-m_Z^2)
 \sin^22\beta+\frac{2b}{\sin2\beta} &\lambda
 v\Lambda\cos2\beta \\
 \lambda
 v(2\mu-\Lambda\sin2\beta)&\lambda
 v\Lambda\cos2\beta & m_{\hat{s}}^2
\end{pmatrix},
\eea
in the basis $(\hat h,\hat H,\hat s)$ defined by
\bea
\hat h &=& \sqrt2 \Big(
({\rm Re}H^0_d - v \cos\beta)\cos\beta
+ ({\rm Re}H^0_u - v \sin\beta)\sin\beta \Big),
\nonumber \\
\hat H &=& \sqrt2 \Big(
({\rm Re}H^0_d - v \cos\beta)\sin\beta
- ({\rm Re}H^0_u - v \sin\beta)\cos\beta \Big),
\nonumber \\
\hat s &=& \sqrt2 \Big(
{\rm Re}S - \langle|S|\rangle \Big),
\eea
with $\langle |H^0_u| \rangle=v \sin\beta$ and $\langle |H^0_d| \rangle = v \cos\beta$
for $v\simeq 174$~GeV.
Here the effective higgsino mass parameter $\mu$, the Higgs scalar $b$-term, and the mixing
parameter $\Lambda$ are determined by
\bea
\mu &=& \lambda \langle S \rangle,
\nonumber \\
b &=& A_\lambda \lambda \langle S \rangle + \lambda \langle \partial_S f \rangle^*,
\nonumber \\
\Lambda &=& A_\lambda - \langle \partial^2_S f \rangle^*.
\eea
If there is no mixing, $\hat h$ acts exactly like the SM Higgs boson with a mass determined
by $m_0$ and $\lambda$.
Including radiative corrections, which mainly come from top and stop loops
\cite{Okada:1990vk}, $m_0$ reads
\bea
m^2_0 = m^2_Z + \frac{3m^4_t}{4\pi^2 v^2}
\ln\left(\frac{m^2_{\tilde t}}{m^2_t}\right)
+ \frac{3m^4_t}{4\pi^2 v^2} \left( X^2_t - \frac{1}{12} X^4_t \right) + \cdots,
\eea
where $m_{\tilde t}$ is the stop mass, and $X_t=(A_t-\mu\cot\beta)/m_{\tilde t}$ is the stop
mixing parameter.
Note that $m_0$ determines how heavy the SM-like Higgs boson can be within the MSSM,
as it basically corresponds to the mass at large $\tan\beta$ in the decoupling limit
of MSSM.\footnote{
There is an additional contribution from Higgs-singlino-higgsino loops, which
is insensitive to $\tan\beta$, and can increase $m_0$ by a few GeV if both the singlino
and higgsino are around the weak scale \cite{Jeong:2012ma,Nakayama:2011iv}.
}
There can be sizable radiative corrections also to other elements in the mass matrix,
which lead to shifts of $b$, $\Lambda$, and $m_{\hat s}$.
The mass eigenstates are found by diagonalizing the mass matrix (\ref{mass-squared}),
which introduces three mixing angles, $\theta_i$:
\bea
 \begin{pmatrix}
  h\\
  H\\
  s
 \end{pmatrix}
 \equiv
\begin{pmatrix}
 &&\\
 &O_{\alpha\hat\beta} & \\
 & &
\end{pmatrix}
\begin{pmatrix}
 \hat{h} \\
 \hat{H} \\
 \hat{s}
\end{pmatrix}
=
\begin{pmatrix}
 \,c_1c_2\, & \,-s_1\, & \,-c_1s_2\, \\
 \,c_2c_3s_1-s_2s_3\, & \,c_1c_3\, & \,-c_3s_1s_2-c_2s_3\, \\
 \,c_3s_2+c_2s_1s_3\, & \,c_1s_3\, & \,c_2c_3-s_1s_2s_3\,
\end{pmatrix}
\begin{pmatrix}
 \hat{h}\\
 \hat{H}\\
 \hat{s}
\end{pmatrix},
\eea
for $c_i=\cos\theta_i$ and $s_i=\sin\theta_i$, where $\alpha=\{h,H,s\}$
and $i=\{1,2,3\}$.
The Lagrangian parameters are written in terms of mass eigenvalues
and mixing angles \cite{diphoton-NMSSM}.
Particularly important are the relation for $m_0$,
\bea
\label{NMSSM-relations-m0}
m^2_0 =
m^2_h
+ (m^2_H - m^2_h) O_{H\hat h}(O_{H\hat h}+O_{H\hat H}\tan2\beta)
- (m^2_h-m^2_s) O_{s\hat h}(O_{s\hat h}+O_{s\hat H}\tan2\beta),
\eea
and the relations for $\lambda$ and $\mu$,
\bea
\label{NMSSM-relations-lam2}
\hspace{-0.5cm}
\lambda^2 v^2 &=& m^2_Z - \frac{2}{\sin4\beta}
\left(
(m^2_H - m^2_h) O_{H\hat h} O_{H\hat H}
- (m^2_h-m^2_s)O_{s\hat h} O_{s\hat H} \right),
\\
\label{NMSSM-relations-lamMu}
\hspace{-0.5cm}
\lambda v \mu &=&
\frac{1}{2} (m^2_H-m^2_h) O_{H \hat s}(O_{H \hat h}+O_{H\hat H}\tan2\beta)
- \frac{1}{2} (m^2_h-m^2_s) O_{s\hat s}(O_{s\hat h}+O_{s\hat H}\tan2\beta).
\eea
These relations allow us to translate the constraints on $m_0$, $\lambda$, and $\mu$
into the constraints on the mass eigenvalues $m_\alpha$ and the mixing angles $\theta_i$,
and vice versa.

\subsection{SM-like Higgs boson}

We identify $h$ as the scalar particle discovered at the LHC since it has properties
close to those of the SM Higgs boson for small mixing.
It interacts with SM particles via
\bea
{\cal L}_{\rm eff} &=& C_V \frac{\sqrt2 m^2_W}{v} h W^+_\mu W^-_\mu
+ C_V \frac{m^2_Z}{\sqrt2 v}h Z_\mu Z_\mu
- C_f \frac{m_f}{\sqrt2 v} h \bar f f
\nonumber \\
&&
+\, C_g \frac{\alpha_s}{12\sqrt2 \pi v} h G^a_{\mu\nu} G^a_{\mu\nu}
+ C_\gamma \frac{\alpha}{\sqrt2 \pi v} h A_{\mu\nu}A_{\mu\nu},
\eea
around the weak scale.
Here the Higgs couplings to the vector bosons and the SM fermions $f$ read
\bea
\label{Higgs-coupling-1}
C_V = c_1 c_2, \quad
C_t = c_1 c_2 + s_1 \cot\beta, \quad
C_b = C_\tau = c_1 c_2 - s_1 \tan\beta,
\eea
at tree-level, and so $h$ has $C_V=C_f=1$ in the limit of vanishing mixing angles.
On the other hand, the couplings to massless gluons and photons are radiatively induced
mainly from the $W$-boson and top-quark loops,
\bea
\label{Higgs-coupling-2}
C_g &\simeq& 1.03 C_t - 0.06 C_b + \delta C_g,
\nonumber \\
C_\gamma &\simeq& 0.23 C_t - 1.04 C_V + \delta C_\gamma,
\eea
where $\delta C_g$ and $\delta C_\gamma$ are the contributions from superparticle loops.
The SUSY contribution $\delta C_g$ can be sizable if the stops are relatively light,
and is approximately estimated as
\bea
\label{delta-Cg}
\delta C_g \approx \frac{1}{4}(2-X^2_t)\frac{m^2_t}{m^2_{\tilde t}} C_t + \cdots,
\eea
for small mass splitting between the two stops \cite{Arvanitaki:2012,Blum:2013}.
In the presence of scalar mixing, the Higgs coupling to photons receives a contribution
from the chargino loops \cite{Carmi:2012in},
\bea
\label{delta-Cr}
\delta C_\gamma
\approx -0.17 \frac{\lambda v}{|\mu|} \cos\theta_1 \sin\theta_2
+ \frac{1}{18}(2-X^2_t)\frac{m^2_t}{m^2_{\tilde t}} C_t
+ \cdots,
\eea
assuming small mixing between the charged wino and higgsinos for simplicity.
Here the first term comes from the charged-higgsino loops, and it either enhances or reduces
the Higgs coupling to photons depending on the singlet-doublet mixing $\theta_2$.
The second term is the contribution from the stop loops, and the ellipsis includes
other SUSY contributions, which are small unless one considers large left-right mixing
of the third generation sfermions, or a small mass around the weak scale for
the charged Higgs boson.

The Higgs sector is constrained by the Higgs boson data from the LHC experiments.
The signal rate of $h$ at the LHC can be estimated in terms of the effective Higgs couplings
by using the well-known production and decay properties of the SM Higgs boson
\cite{Djouadi:2005gi}.
The signal strength normalized by the SM value is given by
\bea
R^{\rm incl}_{VV}
&=&
\frac{\sigma(pp\to h)}{\sigma(pp\to h)|_{\rm SM}}
\times
\frac{{\rm Br}(h\to VV)}{{\rm Br}(h\to VV)|_{\rm SM}}
\nonumber \\
&\simeq&
\frac{(0.92 C^2_g + 0.12 C^2_V)C^2_V}{0.62 C^2_b + 0.26 C^2_V + 0.12 C^2_t},
\eea
for the inclusive $WW/ZZ$ channel, where we have assumed that the Higgs decay rate
into non-SM particles is negligible.
For other channels, one finds
\bea
R^{\rm incl}_{bb} &=& R^{\rm incl}_{\tau\tau} = \frac{C^2_b}{C^2_V} R^{\rm incl}_{VV},
\nonumber \\
R^{\rm incl}_{\gamma\gamma} &\simeq& \frac{1.49 C^2_\gamma}{C^2_V} R^{\rm incl}_{VV}.
\eea
As it should be, the NMSSM leads to $R^{\rm incl}_{xx}=1$ for each channel in the limit
that the mixing angles vanish and the superparticles are decoupled with heavy masses,
i.e.~for $\theta_i=0$ and $\delta C_g=\delta C_\gamma=0$.

\begin{figure}[t]
\begin{center}
\begin{minipage}{16.4cm}
\centerline{
{\hspace*{0cm}\epsfig{figure=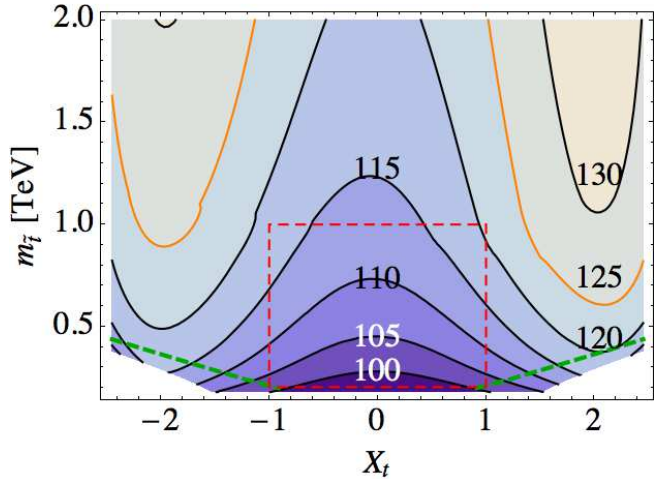,angle=0,width=8.6cm}}
}
\caption{
The dependence of $m_0$ on the stop mass and mixing parameter
in the large $\tan\beta$ regime.
In the MSSM $m_0$ sets the upper limit of the SM-like Higgs boson mass.
We plot the constant contours of $m_0=105,\,110,\,115,\,120,\,\cdots,$~GeV,
respectively, by taking $\mu=200$~GeV and $\tan\beta=40$.
The lightest stop has a mass smaller than $200$~GeV in the region below the
dashed green line.
The region of our interest is inside and around the red dashed box.
}
\label{fig:Higgs-mass}
\end{minipage}
\end{center}
\end{figure}

Another important constraint comes from the observed Higgs boson mass, $m_h\simeq 126$~GeV.
In the MSSM, the upper limit of $m_h$ is set by $m_0$.
Fig.~\ref{fig:Higgs-mass} shows the dependence of $m_0$ on the stop mass and mixing parameters,
%
which is obtained using {\tt FeynHiggs 2.10.0} \cite{Heinemeyer:1998yj,Degrassi:2002fi}.
To get $m_h\simeq 126$~GeV within the MSSM, one needs heavy stops above about $7$~TeV,
or large stop mixing around $X^2_t=6$.
On the other hand, the NMSSM can accommodate a $126$~GeV Higgs boson with SM-like properties
for stops around 1~TeV or even below, without relying on large stop mixing.
This is because $m_h$ can be enhanced
by the additional effects associated with the gauge singlet $S$.
One effect is the tree-level $F$-term contribution, $\Delta m^2 = \lambda^2 v^2 \sin^2 2\beta$,
which becomes sizable at large $\lambda$ and small $\tan\beta$ .
Another effect arises from the mixing with the singlet scalar.

In this paper we focus on the supersymmetric SM with stops around 1~TeV or below,
as would be expected if SUSY is to stabilize the weak scale against large radiative corrections.
The region of our interest is around and inside the red dashed box in Fig.~\ref{fig:Higgs-mass}.
Especially we will focus on the case with $m_s<m_h$ so that the mixing effect raises
the Higgs boson mass $m_h$, and examine how much the Higgs sector is constrained by the current
experimental data and theoretical considerations.

\section{NMSSM with a light singlet scalar}\label{sec:NMSSM-light-singlet}

We first specify the range of $m_0$ and $m_\alpha$ in our scenario with a light singlet
scalar, and then
move on to the experimental constraints on the Higgs sector and their implications.
The case of our interest is that the CP-even neutral Higgs sector has the mass spectrum
\bea
\frac{1}{2}m_h < m_s < m_h,
\eea
with $m_h\simeq 126$~GeV, while the superparticles including the stops are
around or below TeV.
Here we do not assume a particular mass spectrum for the superparticles or
particular mediation mechanism of SUSY breaking.
The lower bound on $m_s$ is to avoid the process $h\to ss$.
For the heavier Higgs boson $H$, we consider
\bea
350\,{\rm GeV} \lesssim m_H \lesssim 1\,{\rm TeV},
\eea
taking into account the experimental constraint from $b\to s\gamma$, which requires
the charged Higgs scalar to be heavier than about $350$~GeV barring cancellation with
other superparticle contributions \cite{Gambino:2001ew}.

The NMSSM can explain the observed Higgs boson mass even for $m_0$ around or below
$120$~GeV,
i.e.~for stops having $m_{\tilde t}\lesssim 1$~TeV and $X^2_t\lesssim 1$,
because there are extra contributions associated with the singlet scalar.
To be specific, we take
\bea
100\,{\rm GeV} \lesssim m_0 \lesssim 120\,{\rm GeV},
\eea
keeping in mind that $m_0$ sets the upper limit of the SM-like Higgs boson mass
in the MSSM, and has a dependence on the stop mass and mixing as plotted
in Fig.~\ref{fig:Higgs-mass}.
For the coupling $\lambda$, we impose
\bea
\label{lambda-bound}
0.01< \lambda < 1,
\eea
at the weak scale.
The upper bound is to ensure that the model remains perturbative up to the conventional
GUT scale.
This requires $\lambda$ to be smaller than 0.7--0.8 at the weak scale, which is slightly
relaxed in the presence of extra heavy particles charged under the SM gauge group
\cite{Masip:1998jc,Barbieri:2007tu}.
The perturbative bound can be relaxed further in U$(1)$ gauge extensions \cite{Kyae:2012df}
or in extensions with hidden gauge sector coupled to $S$ \cite{Nakayama:2011iv}.
On the other hand, the LEP constraint on the chargino mass requires $\mu$
larger than about $100$~GeV \cite{PDG}, implying that the singlet scalar has a VEV above
$\lambda^{-1} \times 100$~GeV.
We have put a mild lower bound on $\lambda$ following that the singlet scalar would not
have a VEV much larger than TeV in low scale SUSY.

Finally it should be noted that the relation (\ref{NMSSM-relations-m0}) and
(\ref{NMSSM-relations-lam2}) lead to
\bea
\lambda^2 v^2 = m^2_Z + \frac{1}{\sin^2 2\beta}\left(
(m^2_h-m^2_0) - (m^2_h-m^2_s) O^2_{s\hat h}
+ (m^2_H-m^2_h)O^2_{H\hat h} \right).
\eea
Thus, for $m_s<m_h$, small $\lambda$ requires sizable $O^2_{s\hat h}$ and small
$O^2_{H\hat h}$.
This simply reflects the fact that $m_h$ receives a positive contribution both from
the tree-level $F$-term potential associated with the singlet and the singlet-doublet
mixing effect, whereas a negative contribution from the doublet-doublet mixing effect.

\subsection{LHC constraints on Higgs mixing}

The Higgs couplings to SM fermions and vector bosons (\ref{Higgs-coupling-1}) are fixed by
$\theta_1$, $\theta_2$, and $\tan\beta$,
while the couplings to gluons and photons (\ref{Higgs-coupling-2}) can receive an
additional sizable contribution from superparticle loops.
This implies that the signal rate of the SM-like Higgs boson is a function written
\bea
R_{xx} =
R_{xx}(\theta_1,\theta_2,\tan\beta,\delta C_g,\delta C_\gamma),
\eea
for each decay channel, $h\to xx$, with $x=\{W,\,Z,\,b,\,\tau,\,\gamma\}$.
Hence the Higgs signal strength measured at the LHC puts a constraint on the mixing angle
$\theta_1$ and $\theta_2$.
To examine the constraint, one also needs to estimate $\delta C_g$ and $\delta C_\gamma$.
The stop searches at the LHC suggest that stop mass lighter than about $600$~GeV is
excluded depending on the mass of the lightest neutralino if the stop is kinematically
allowed to decay into the top quark and the lightest neutralino \cite{ATLAS-stop,CMS-stop}.
Taking this into account, we separate two cases according to the stop mass, and find
$m_0$ and $\delta C_g$ to be
\bea
\label{cases-stop-mass}
\hspace{-0.5cm} && 600{\rm GeV} \lesssim m_{\tilde t} \lesssim 1{\rm TeV}
\,\quad\,:\,
105{\rm GeV} \lesssim m_0 \lesssim 120{\rm GeV},
\,\, \delta C_g \lesssim 0.04 C_t,
\nonumber \\
\hspace{-0.5cm}
&& 200{\rm GeV} \lesssim m_{\tilde t} \lesssim 600{\rm GeV}
\,:\,
100{\rm GeV} \lesssim m_0 \lesssim 115{\rm GeV},
\,\, \delta C_g \lesssim 0.21 C_t,
\eea
taking $X^2_t\lesssim 1$.
On the other hand, the Higgs coupling to photons receives SUSY contributions,
\bea
|\delta C_\gamma| \lesssim 0.3\,|\cos\theta_1 \sin\theta_2|
+ \frac{2}{9}\delta C_g,
\eea
from the chargino and stop loops, where we have used the relation (\ref{delta-Cr}) taking
$\lambda<1$ as limited by the perturbativity constraint, and $|\mu|>100$~GeV
considering the LEP bound on the chargino mass.

Let us examine the range of Higgs mixing compatible with the current experimental data
reported by the ATLAS \cite{ATLAS-Higgs-data,ATLAS-Higgs-data-update} and
CMS \cite{CMS-Higgs-data,CMS-Higgs-data-update}
collaborations, respectively.
For the decay modes $h\to WW^*$ and $ZZ^*$, we consider the signal rate in the inclusive
channel assuming that it is dominated by the gluon-gluon fusion (ggF) production.
For the fermionic modes, we focus on the vector boson fusion (VBF) and
vector boson associated (VH) production.
On the other hand, the signal rate in the diphoton mode requires a more careful treatment
because the reported data suggests a correlation between the diphoton rates in the ggF
and VBF cannel.
To employ a $\chi^2$ analysis, we define the measures
\bea
R^X_{\gamma\gamma} &=& 1 + (R^{\rm ggF/ttH}_{\gamma\gamma}-1)\cos\varphi +
(R^{\rm VH/VBF}_{\gamma\gamma}-1)\sin\varphi,
\nonumber \\
R^Y_{\gamma\gamma} &=& 1 - (R^{\rm ggF/ttH}_{\gamma\gamma}-1)\sin\varphi +
(R^{\rm VH/VBF}_{\gamma\gamma}-1)\cos\varphi,
\eea
and take $\cos\varphi=0.98$ for the ATLAS data and $0.97$ for the CMS data so that
$R^X_{\gamma\gamma}$ and $R^Y_{\gamma\gamma}$ can be treated as independent.
Table \ref{signaltable} summarizes the Higgs signal rates we will use in the analysis.
\begin{table}[t]
 \begin{tabular}{|c||c|c|c|c|c|c|}
  \noalign{\vspace{1mm}}
  \hline
   & \,\,$R^{\rm incl}_{WW}$\,\, & \,\,$R^{\rm incl}_{ZZ}$\,\,
   & \,\,$R^{\rm VH/VBF}_{bb}$\,\, & \,\,$R^{\rm VH/VBF}_{\tau\tau}$\,\,
   & \,\,$R^{X}_{\gamma\gamma}$\,\, & \,\,$R^{Y}_{\gamma\gamma}$\,\, \\
  \hline
  \,\,ATLAS\,\, & \,$0.99^{+0.31}_{-0.28}$\,
  & \,$1.43^{+0.40}_{-0.35}$\, & \multicolumn{2}{|c|}{\,$1.09^{+0.36\ast}_{-0.32}$\,}
  & \,$ 1.49 \pm 0.36 $\, & \,$ 0.61 \pm 0.75 $\,
  \\
  \hline
  \,\,CMS\,\, & \,$0.68\pm 0.20$\, & \,$0.92\pm 0.28$\,
  & \,$1.15\pm 0.62$\, & \,$1.10\pm 0.41$\,
  & \,$ 1.42 \pm 0.31 $\, & \,$ 0.89 \pm 0.61 $\,
  \\
  \hline
  \noalign{\vspace{2mm}}
 \end{tabular}
\caption{The summary of the Higgs signal rates. These are evaluated at
 $m_h=125.5$~GeV for the ATLAS while $m_h=125.7$~GeV for the CMS. The number
with asterisk is taken from Ref.~\cite{Aad:2013wqa}, and the others from
Refs.~\cite{ATLAS-Higgs-data,ATLAS-Higgs-data-update,CMS-Higgs-data,CMS-Higgs-data-update}.}
\label{signaltable}
\end{table}
Note that the signal rate normalized by the SM prediction is given by
\bea
R^{\rm VH/VBF}_{bb} &\simeq& \frac{C^2_VC^2_b}{0.62C^2_b+0.26C^2_V+0.12C^2_t},
\nonumber \\
R^{\rm VH/VBF}_{\gamma\gamma} &\simeq&
\frac{1.52 C^2_VC^2_\gamma}{0.62C^2_b+0.26C^2_V+0.12C^2_t},
\eea
for $m_h\simeq 126$~GeV.
In the analysis we include the SUSY contributions $\delta C_g$ and $\delta C_\gamma$
lying in the range indicated above, and minimize $\chi^2$ at each point on
the $(\theta_1,\theta_2)$ plane assuming a Gaussian distribution.

\begin{figure}[t]
\begin{center}
\begin{minipage}{16.4cm}
\centerline{
{\hspace*{0cm}\epsfig{figure=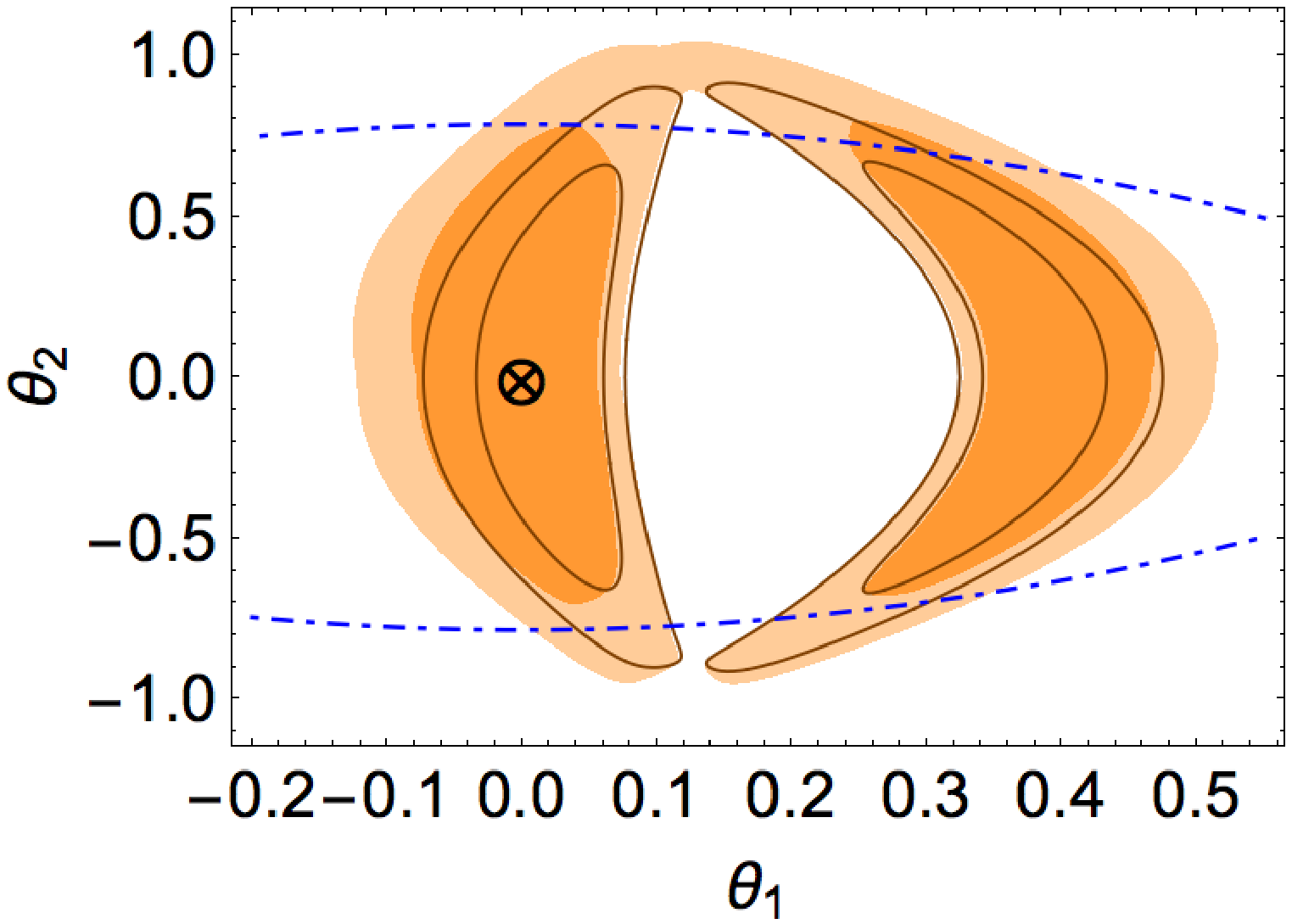,angle=0,width=5.4cm}}
{\hspace*{.2cm}\epsfig{figure=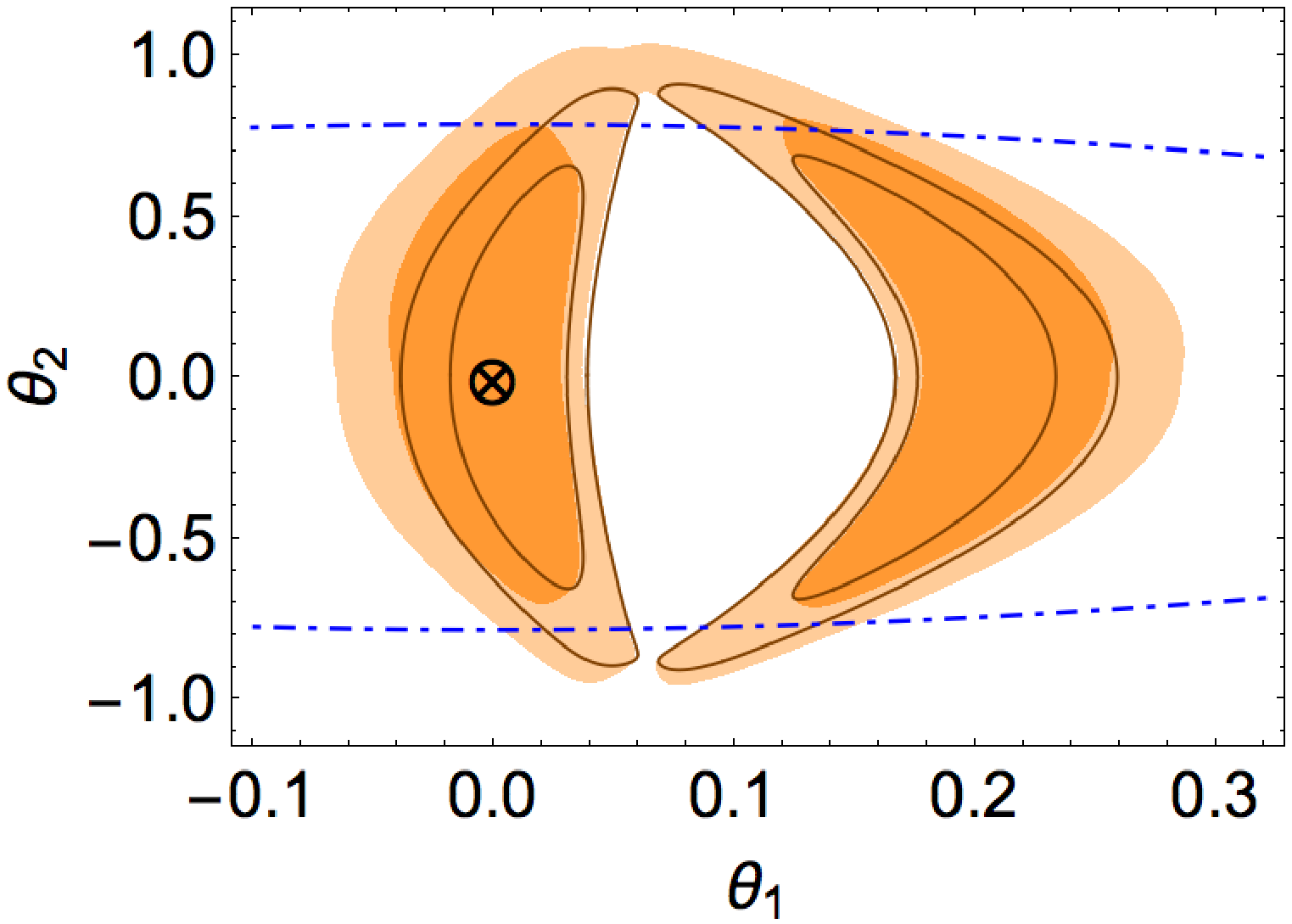,angle=0,width=5.4cm}}
{\hspace*{.2cm}\epsfig{figure=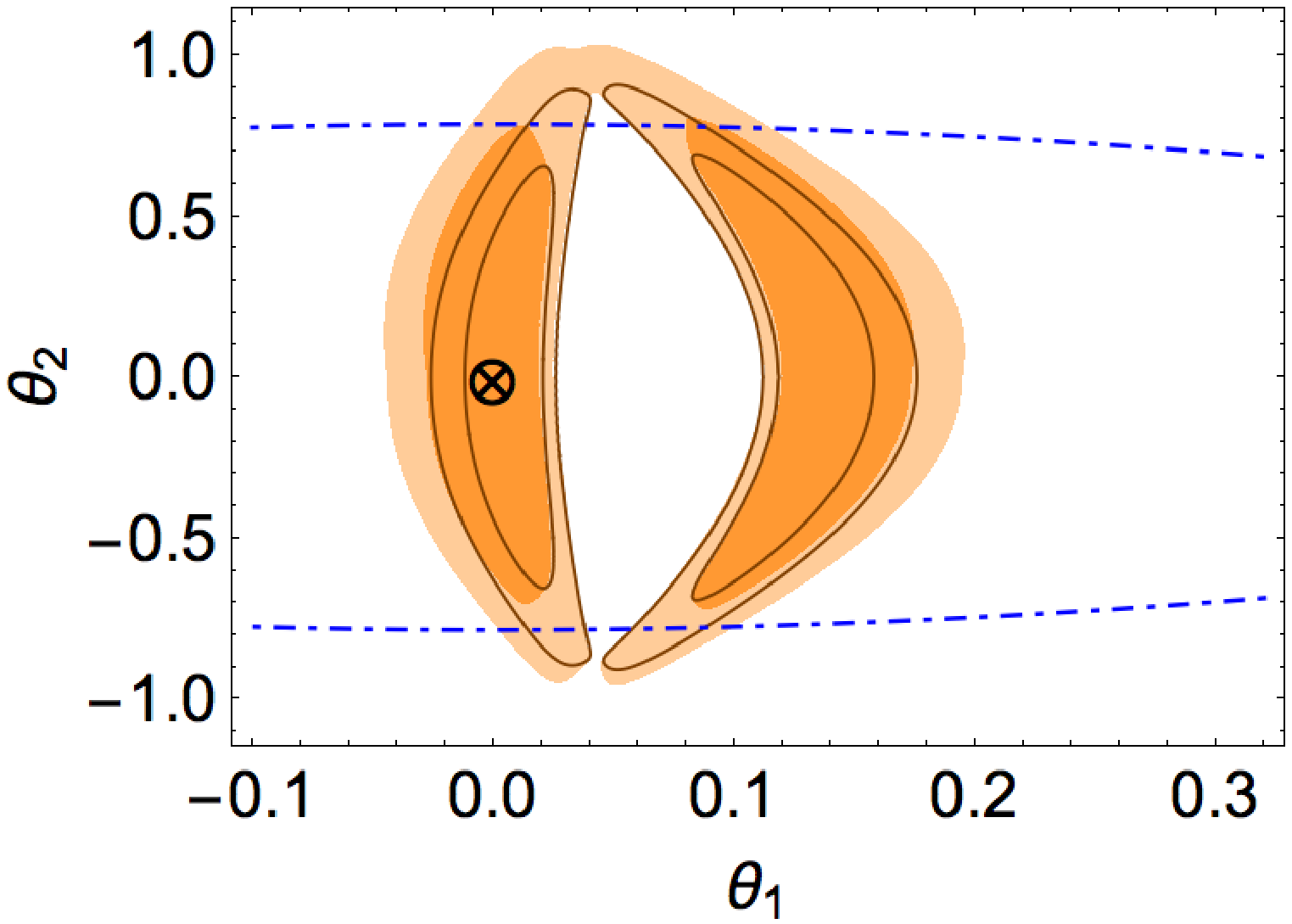,angle=0,width=5.4cm}}
}
\vspace{0.3cm}
\centerline{
{\hspace*{0cm}\epsfig{figure=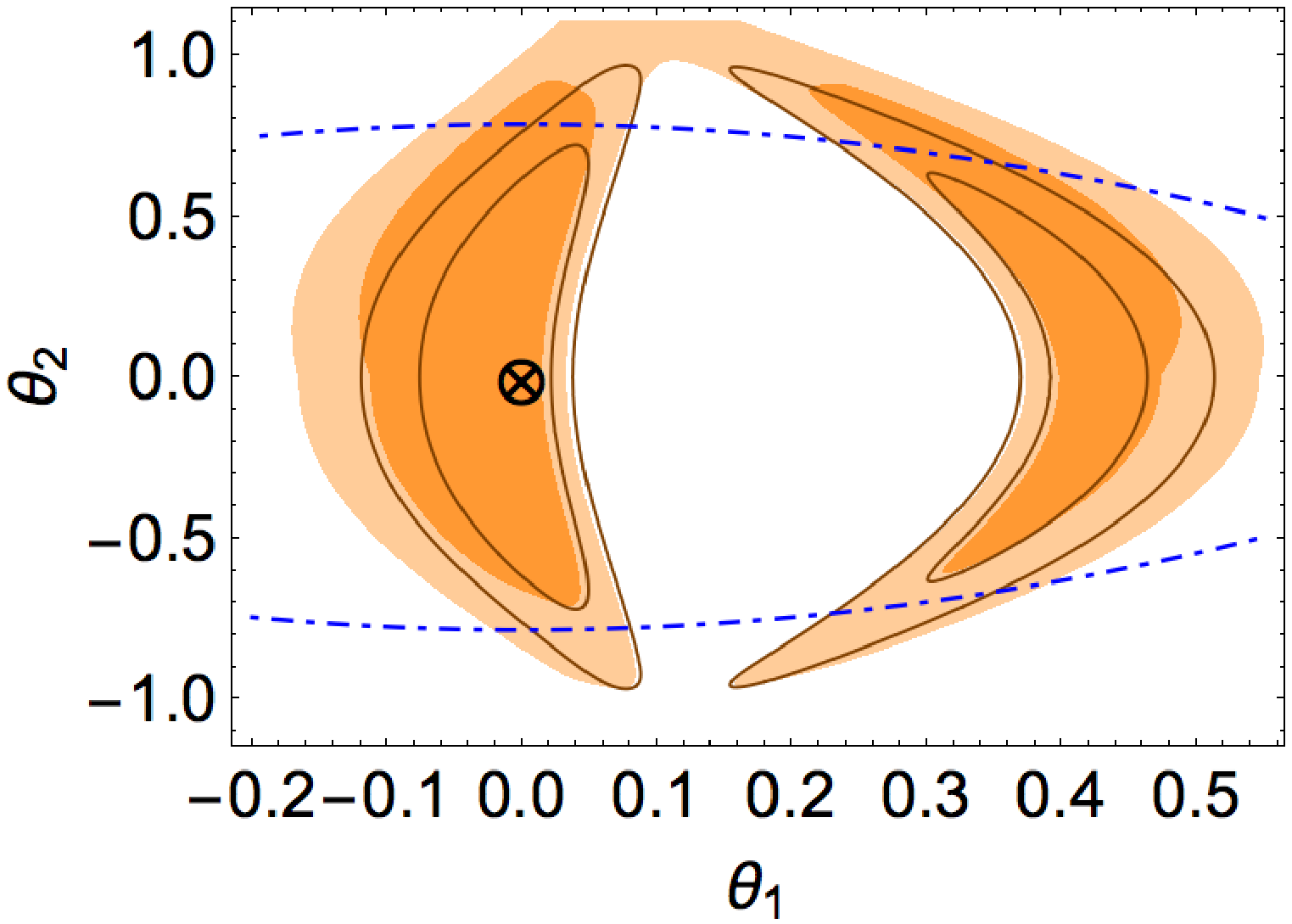,angle=0,width=5.4cm}}
{\hspace*{.2cm}\epsfig{figure=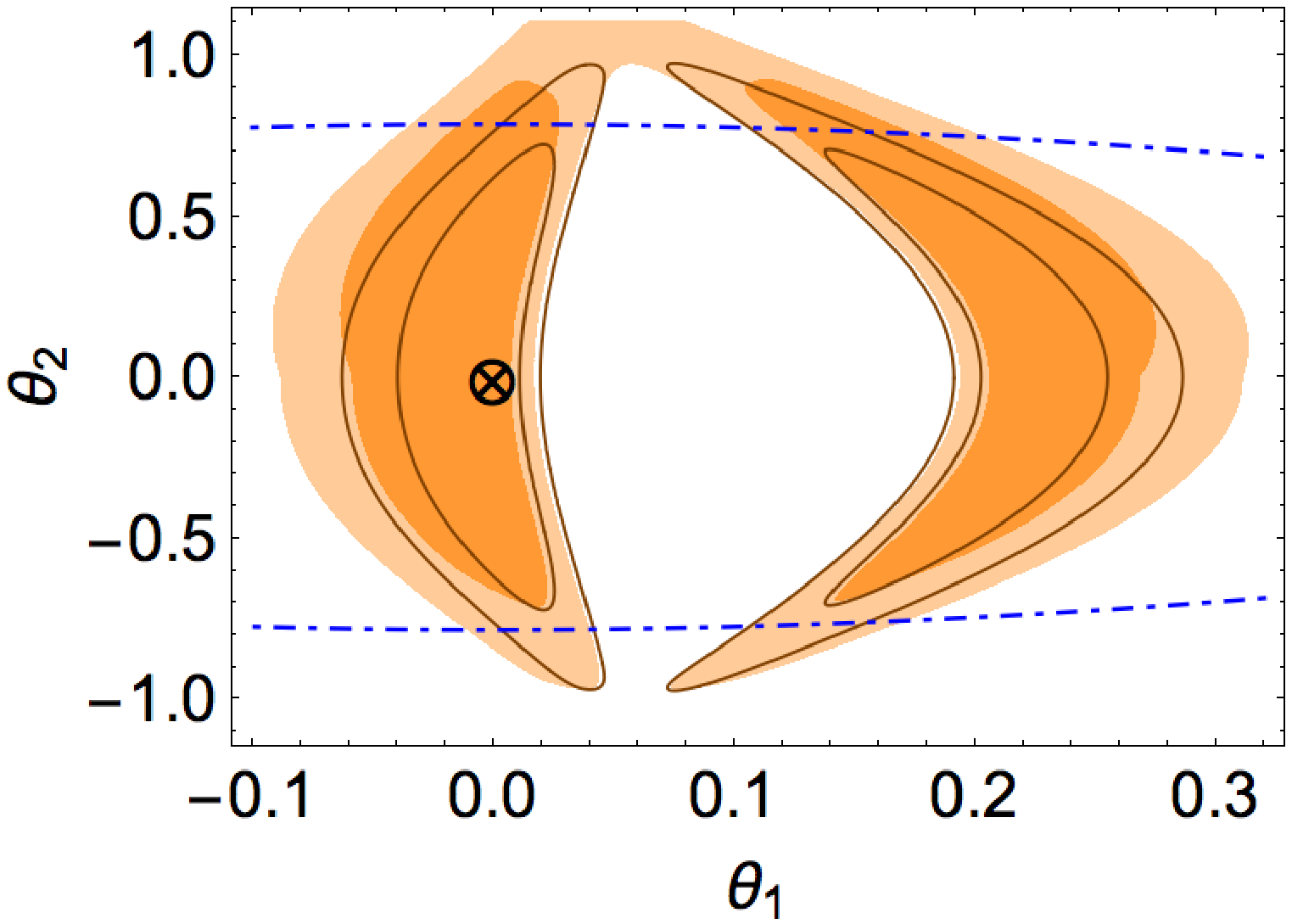,angle=0,width=5.4cm}}
{\hspace*{.2cm}\epsfig{figure=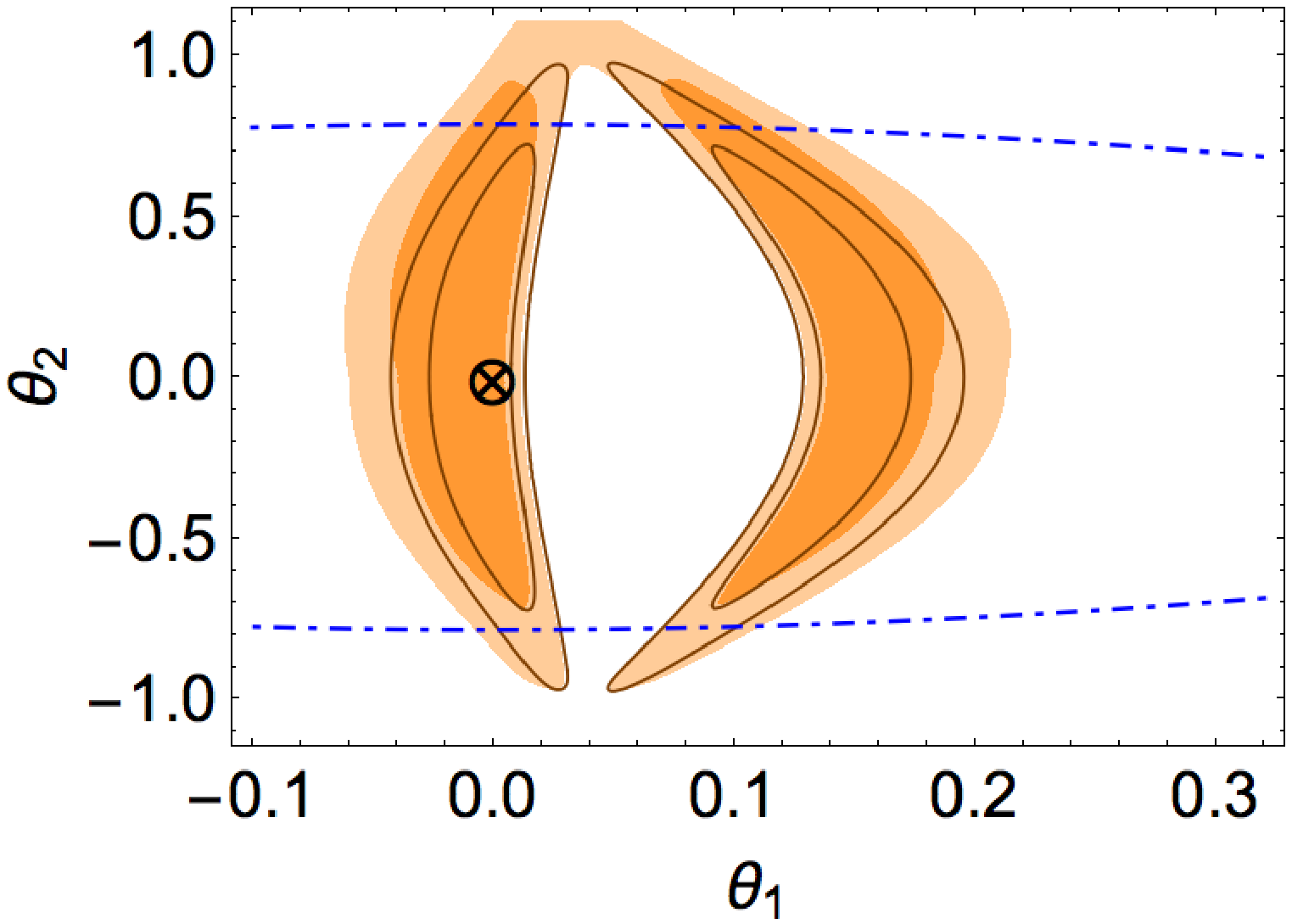,angle=0,width=5.4cm}}
}
\caption{
Higgs mixing compatible with the LHC data.
We show the region of $(\theta_1,\theta_2)$ preferred at the $68$\% (dark orange) and
$95$\% CL (light orange) by the ATLAS (upper) and the CMS (lower) results, respectively.
Here we have taken $m_h=126$~GeV, and $\tan\beta=5,\,10,\,15$ for the left, middle,
right panel.
The $\hat h$ fraction in $h$, which is given by $O^2_{h\hat h}$ and determines
how close $h$ is to the SM Higgs boson, is larger than $0.5$ in the region
between two dot-dashed blue curves.
For comparison, we also plot the $68$\% (outer brown circle) and $95$\% CL
(inner brown circle) preferred region in the limit that the superparticles are
very heavy, i.e.~the case with $\delta C_g=\delta C_\gamma=0$.
}
\label{fig:Higgs-mixing}
\end{minipage}
\end{center}
\end{figure}

Fig.~\ref{fig:Higgs-mixing} illustrates which region of $(\theta_1,\theta_2)$
is compatible with the current LHC data on the Higgs boson.
The $\hat h$ fraction in $h$ is larger than $0.5$ in the region between the two
dot-dashed blue curves, making $h$ SM-like.
The dark and light orange regions are preferred at the $68$\% and $95$\% CL, respectively,
by the ATLAS (upper) and CMS (lower) measurements.
For comparison, we also plot the $68$\% (outer brown circle) and $95$\% CL (inner brown
circle) preferred region for the case with vanishing $\delta C_g$ and $\delta C_\gamma$.
One can see that sizable scalar mixing is compatible with the current LHC data, and
superparticle contributions to $C_g$ and $C_\gamma$ slightly enlarge the allowed region.
In addition there are a couple of things to note.
The shaded region is not symmetric under $\theta_2\to -\theta_2$ because the Higgs coupling
to photons receives a contribution from the chargino loops combined with
the singlet-doublet mixing.
For given $\tan\beta$, there are two ranges of $\theta_1$ where $h$ can describe
the observed data.
One is around $\theta_1=0$, and the other is around $\theta_1=\arctan(2/\tan\beta)$
\cite{diphoton-NMSSM}.
This is understood from the fact that the Higgs decay $h\to b\bar b$ occurs through
the effective coupling $C_b=c_1c_2-s_1\tan\beta$, and it should be the main decay
mode in order to explain the LHC results \cite{LHC-Higgs-couplings}.
Hence one needs either $C_b\sim 1$ or $C_b\sim -1$.
The former is the case in the region around the origin, $(\theta_1,\theta_2)=(0,0)$.
In the latter case, which is obtained in the region with $\tan\theta_1 = 2/\tan\beta$
and small $\theta_2$, the sign of the Higgs coupling to down-type fermions is opposite
to that of the SM Higgs boson, and consequently the bottom and top quark loops give
the same sign of contributions to the Higgs coupling to gluons.\footnote{
See also Ref.~\cite{Ferreira:2014naa} for the discussion on a wrong-sign Yukawa coupling
in two-Higgs-doublet model.
}
Finally we note that the sensitivity of $C_b$ to $\tan\beta$ results in that the preferred
region gets smaller as $\tan\beta$ increase, as can be seen from the figure.
The future run of the LHC and linear collider experiments will help us to clarify
the viable region of $(\theta_1,\theta_2)$ more accurately, and could determine
the sign of Higgs coupling to bottom-quark pairs.

\subsection{Implication of Higgs mixing on higgsino properties}

\begin{figure}[t]
\begin{center}
\begin{minipage}{16.4cm}
\centerline{
{\hspace*{0cm}\epsfig{figure=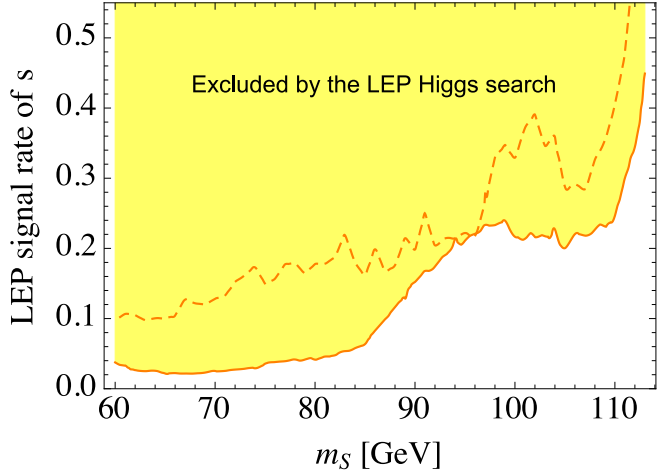,angle=0,width=8.6cm}}
}
\caption{
LEP constraints on the singlet-like Higgs boson $s$.
The shaded region is excluded by the LEP result on Higgs searches
in the channel, $e^+e^- \to Z s \to Z b\bar b$.
The dashed orange curve shows the LEP constraint on the production cross
section of hadronically decaying Higgs bosons.
%
}
\label{fig:LEP}
\end{minipage}
\end{center}
\end{figure}

The Higgs sector is further constrained by the observed Higgs boson mass,
and the LEP results on the Higgs search if the singlet-like Higgs boson is
lighter than $114$~GeV \cite{Barate:2003sz}.
Interestingly, combined with the constraints from the measured Higgs signal rate,
these are found to put an upper bound on $\mu$, requiring higgsinos to be relatively light.
The LEP signal rate of $s$ relative to the SM prediction is given by
\bea
R(e^+e^- \to Z s \to Z b\bar b)
= O^2_{s\hat h} \times {\rm Br}(s\to b\bar b).
\eea
The singlet-like Higgs boson $s$ dominantly decays into bottom quarks for
%
$2m_b < m_s<m_h$ in most of the region of mixing angles.
Note however that ${\rm Br}(s\to b\bar b)$ is highly suppressed when the
$sb\bar b$ coupling vanishes, i.e.~for the scalar mixing satisfying
$s_2c_3+s_1c_2s_3 + c_1s_3\tan\beta=0$.
Fig.~\ref{fig:LEP} shows the LEP constraints on the singlet-like Higgs boson.
In the yellow-shaded region, the signal rate of $s$ in the channel
$e^+e^- \to Z s \to Z b\bar b$ is above the LEP bound.
The dashed orange curve is the LEP constraint on the production cross section
of hadronically decaying Higgs bosons~\cite{Searches:2001aa}.

Let us use the relations (\ref{NMSSM-relations-m0}), (\ref{NMSSM-relations-lam2})
and (\ref{NMSSM-relations-lamMu}) to see why light higgsinos are favored in the NMSSM
with a light singlet scalar when $h$ is arranged to have properties consistent with
the observation.
From (\ref{NMSSM-relations-m0}) and (\ref{NMSSM-relations-lam2}), one obtains
\bea
&&
\left(
1 - \frac{\lambda^2v^2-m^2_Z}{m^2_H-m^2_h}\cos^2 2\beta
+ 2 \frac{m^2_h-m^2_s}{m^2_H-m^2_h} O_{s\hat h}O_{s\hat H}\cot2\beta
\right)
(\lambda^2v^2-m^2_Z) \sin^2 2\beta
\nonumber \\
&&
\quad\quad
=\, (m^2_h-m^2_0)
- \left(1- \frac{m^2_h-m^2_s}{m^2_H-m^2_h} O^2_{s\hat H} \right)
(m^2_h-m^2_s) O^2_{s\hat h}.
\eea
Since we consider the case with $m^2_H\gg m^2_h$ and $\lambda< 1$, the above is
approximated as
\bea
\label{lambda-rel1}
(\lambda^2v^2-m^2_Z)\sin^2 2\beta
\simeq
(m^2_h-m^2_0)
- (m^2_h-m^2_s)O^2_{s\hat h},
\eea
for $\tan\beta\ll (m^2_H-m^2_h)/(m^2_h-m^2_s)$.
Hence the value of $\lambda$ is mainly fixed by $\tan\beta$ and the mixing angles through
$O_{s\hat h}=c_3s_2+c_2s_1s_3$, but insensitively to the precise value of $m_H$,
as long as the Higgs sector has $m^2_H\gg m^2_h$.

The Higgs boson $h$ becomes more SM-like when the mixing parameters $O_{h\hat H}$
and $O_{h\hat s}$ get smaller.
Expanded in powers of $O_{h\hat H}$ and $O_{h\hat s}$, the relation
(\ref{NMSSM-relations-lamMu}) is written
\bea
\label{mu-rel1}
\lambda v \mu &\simeq&
\frac{m^2_h-m^2_s}{1+t^2_3} \left(
\frac{\tan\beta}{\tan^2\beta-1} t_3
+  \frac{1}{2} O_{h\hat s} + \frac{t_3}{2} O_{h\hat H}
\right)
\nonumber \\
&&
+\,
\frac{m^2_H-m^2_h}{1+t^2_3}
\left(
\frac{\tan\beta}{\tan^2\beta-1}
- \frac{t_3}{2} O_{h\hat s} + \frac{1}{2} O_{h\hat H} \right)
(t_3+O_{h\hat s}O_{h\hat H}),
\eea
at the leading order.
Here $t_3\equiv \tan\theta_3$, and we have assumed that $t^2_3$ is not much
larger than unity, for which $H$ has a sizable $\hat H$ component.
In addition, the relation (\ref{NMSSM-relations-lam2}) allows us to estimate
$O_{h\hat H}$ in terms of $O_{h\hat s}$ as
\bea
O_{h\hat H} \simeq
\epsilon + t_3 O_{h\hat s},
\eea
for $m^2_H\gg m^2_h$, neglecting terms in higher order in $O_{h\hat s}$.
The parameter $\epsilon$ is given by
\bea
\epsilon \approx
-\frac{2}{\tan\beta} \frac{\lambda^2 v^2-m^2_Z}{m^2_H-m^2_h}(1+t^2_3),
\eea
and so it is much smaller than $1/\tan\beta$.
Plugging the above relation between mixing parameters into (\ref{mu-rel1}),
one arrives at
\bea
\mu \approx
\frac{m^2_h-m^2_s}{2\lambda v}O_{h\hat s}
+ \frac{m^2_H-m^2_s}{\lambda v}
\frac{1}{\tan\beta}
\frac{O_{h\hat s}(O_{h\hat H}-\epsilon)}{O^2_{h\hat s}+(O_{h\hat H}-\epsilon)^2},
\eea
for nonzero $O_{h\hat s}$, as is required to increase $m_h$ via the singlet-doublet
mixing.
Finally the higgsino mass parameter is found to lie in the range
\bea
\label{mu-upper-bound}
|\mu| \lesssim
\frac{m^2_h-m^2_s}{2\lambda v} |O_{h\hat s}|
+ \frac{m^2_H-m^2_s}{\lambda v}
\frac{1}{\tan\beta}
\frac{|O_{h\hat s}(O_{h\hat H}-\epsilon)|}{O^2_{h\hat s}+(O_{h\hat H}-\epsilon)^2},
\eea
with $\lambda$ approximately determined by the relation (\ref{lambda-rel1}),
\bea
\lambda^2 \approx
\frac{m^2_Z}{v^2}
+ \frac{\tan^2\beta}{4}\left(
\frac{m^2_h-m^2_0}{v^2}
- \frac{m^2_h-m^2_s}{v^2}
\frac{
(O^2_{h\hat s}+(O_{h\hat H}-\epsilon)O_{h\hat H})^2
}{O^2_{h\hat s}+(O_{h\hat H}-\epsilon)^2}
\right).
\eea
Therefore there is an upper limit on $\mu$, depending on how close $h$ is to
the SM Higgs boson.
Note that $\mu$ takes the maximum value when the mixing parameter $O_{h\hat H}$ has
a value, $O^2_{h\hat H}\simeq O^2_{h\hat s}$, for sizable $O_{h\hat s}$
compatible with the LEP constraint.
Here we have used that $\epsilon$ has a tiny value for $m^2_H\gg m^2_h$
and $\lambda<1$.
Another important feature is that the upper bound on $\mu$ grows as $m_s$ decreases,
because the right hand side of (\ref{mu-upper-bound}) increases while $\lambda$
decreases.
However the LEP constraints on the singlet-doublet mixing become stringent when
the singlet-like Higgs boson is light.

\begin{figure}[t]
\begin{center}
\begin{minipage}{16.4cm}
\centerline{
{\hspace*{0cm}\epsfig{figure=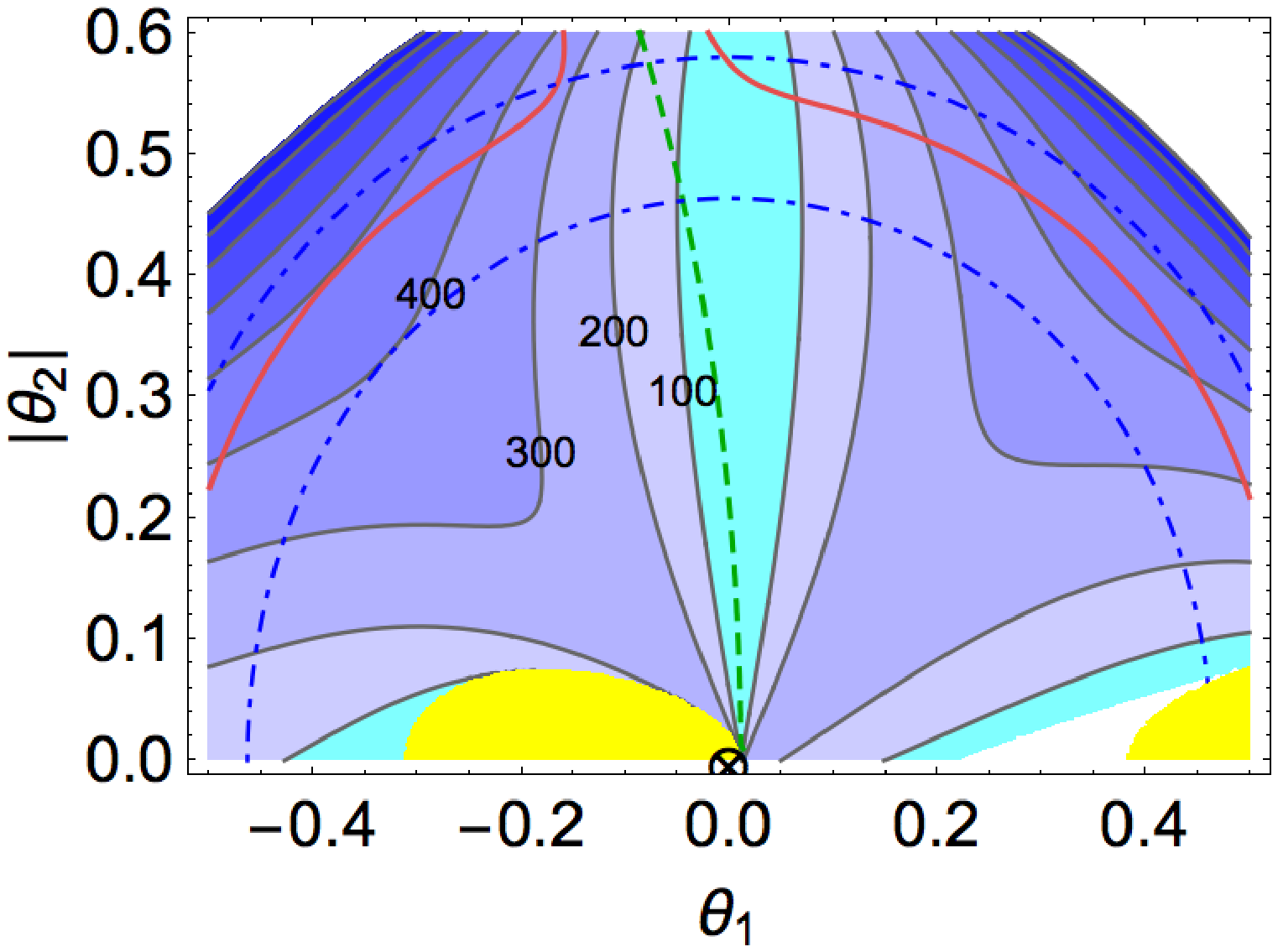,angle=0,width=7.8cm}}
{\hspace*{.4cm}\epsfig{figure=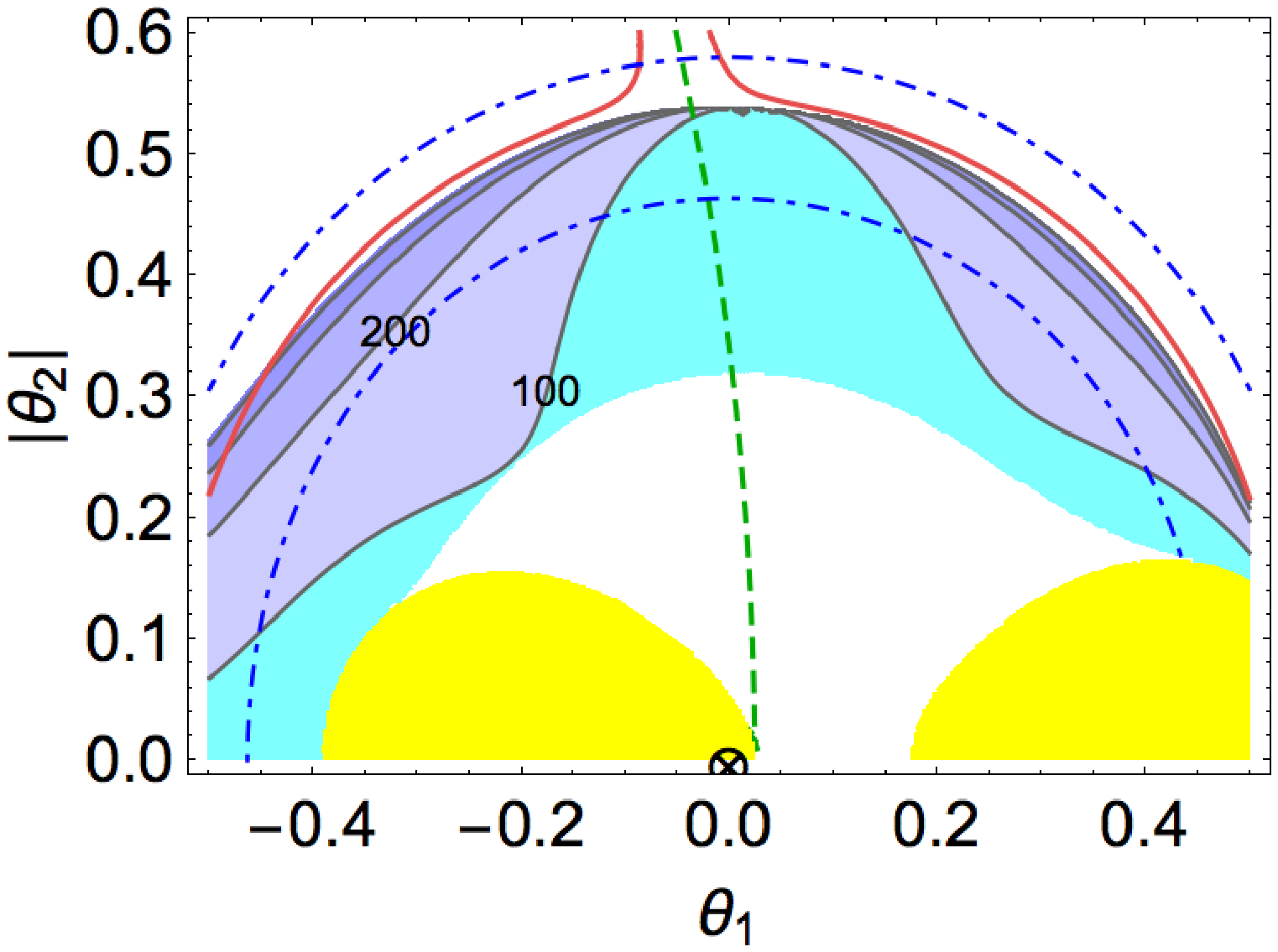,angle=0,width=7.8cm}}
}
\caption{
Higgsino mass parameter $\mu$ for $\tan\beta=5$ (left) and $\tan\beta=10$ (right)
in the NMSSM with $m_0=120$~GeV, $m_h=126$~GeV, $m_H=600$~GeV and $m_s=95$~GeV.
On the $(\theta_1,\theta_2)$ plane, we plot the constant contours of
$|\mu|=100,\,200,\,300,\,400,\,\cdots,$~GeV, respectively, with darker blue indicating
larger $\mu$.
The cyan-shaded region gives $\mu$ smaller than $100$~GeV, and the unshaded region
leads to $\lambda>1$.
The yellow-shaded region is excluded because the relation (\ref{NMSSM-relations-m0})
cannot be satisfied for a real value of $\theta_3$,
while the region outside the red curve is excluded by the LEP constraints on
the singlet-like Higgs boson $s$.
The scalar mixing makes the $sb\bar b$ coupling vanish at tree level along the dashed green
line, relaxing the LEP constraint on the signal rate of $s$.
We also show the $\hat h$ fraction in $h$:
$O^2_{h\hat h}=0.8$ ($0.7$) on the dot-dashed blue half-circle of smaller (larger) radius.
}
\label{fig:mu}
\end{minipage}
\end{center}
\end{figure}

We are ready to analyze how strongly the higgsino mass parameter is constrained in
the NMSSM with a light singlet scalar.
Our strategy is to examine the value of $\mu$ on the $(\theta_1,\theta_2)$ plane
for fixed $m_0$, $m_\alpha$, and $\tan\beta$.
Then $\theta_3$ is determined by (\ref{NMSSM-relations-m0}), and subsequently one
can compute $\mu$ and $\lambda$ using (\ref{NMSSM-relations-lam2}) and
(\ref{NMSSM-relations-lamMu}).
Here we notice that there exist at most two values of $\theta_3$ satisfying the
relation (\ref{NMSSM-relations-m0}).
If there are two solutions at a given point on the $(\theta_1,\theta_2)$ plane, we will
take the value of $\theta_3$ that gives larger $\mu$.
Fig.~\ref{fig:mu} shows the higgsino mass parameter $|\mu|$ for
$\tan\beta=5$ (left) and $\tan\beta=10$ (right) in the NMSSM with $m_0=120$~GeV,
$m_h=126$~GeV, $m_H=600$~GeV and $m_s=95$~GeV.
For $m_s=95$~GeV, the LEP constraint
\bea
O^2_{s\hat h} = (s_1c_2c_3-s_2s_3)^2 \,\lesssim\, \frac{0.24}{{\rm Br}(s\to b\bar b)}
\eea
is satisfied in the region inside the thick red curve.
Note that along the dashed red line $s$ does not couple to the bottom quark at tree level,
and thus the decay rate $s\to b\bar b$ is highly suppressed.
As discussed above, large $\mu$ favors small $m_s$, but the LEP constraint becomes stronger
as $s$ gets lighter.
The contours of $|\mu|=100,\,200,\,300,\,400,\,\cdots,$~GeV are shown by the solid gray
lines, with darker blue indicating larger $\mu$.
The cyan-shaded region gives $\mu$ smaller than $100$~GeV and so is in conflict with
the LEP constraint on the chargino mass, while the yellow-shaded region is excluded since
the relation (\ref{NMSSM-relations-m0}) has no solution for real $\theta_3$.
We also show the $\hat h$ fraction in the SM-like Higgs boson $h$: $O^2_{h\hat h}=0.8$ (0.7)
on the dot-dashed blue half-circle of smaller (larger) radius.
As can be seen from the figure, the higgsino mass is required to be small as $h$ becomes
more SM-like, and $\tan\beta$ increases.
In the unshaded region, $\lambda$ is above the perturbative bound, requiring new physics
below the GUT scale.
If one allows $\lambda>1$ at the weak scale, there appears additionally a viable region
with $|\mu|>100$~GeV,
but only in the outer unshaded region where $h$ is less SM-like.
Notice that the value of $\theta_3$ is fixed by solving the relation (\ref{NMSSM-relations-m0}),
and there are two solutions in the area except the yellow-shaded region.
Inserting the two solutions into (\ref{NMSSM-relations-lam2}), one finds that the value of
$\lambda^2$ either keeps growing or changes sign when one crosses the boundary between
the blue-shaded and outer unshaded region.
Combined with the relation (\ref{NMSSM-relations-lamMu}), this explains why $\mu$
is large near the boundary.

\begin{figure}[t]
\begin{center}
\begin{minipage}{16.4cm}
%
\centerline{
{\hspace*{0cm}\epsfig{figure=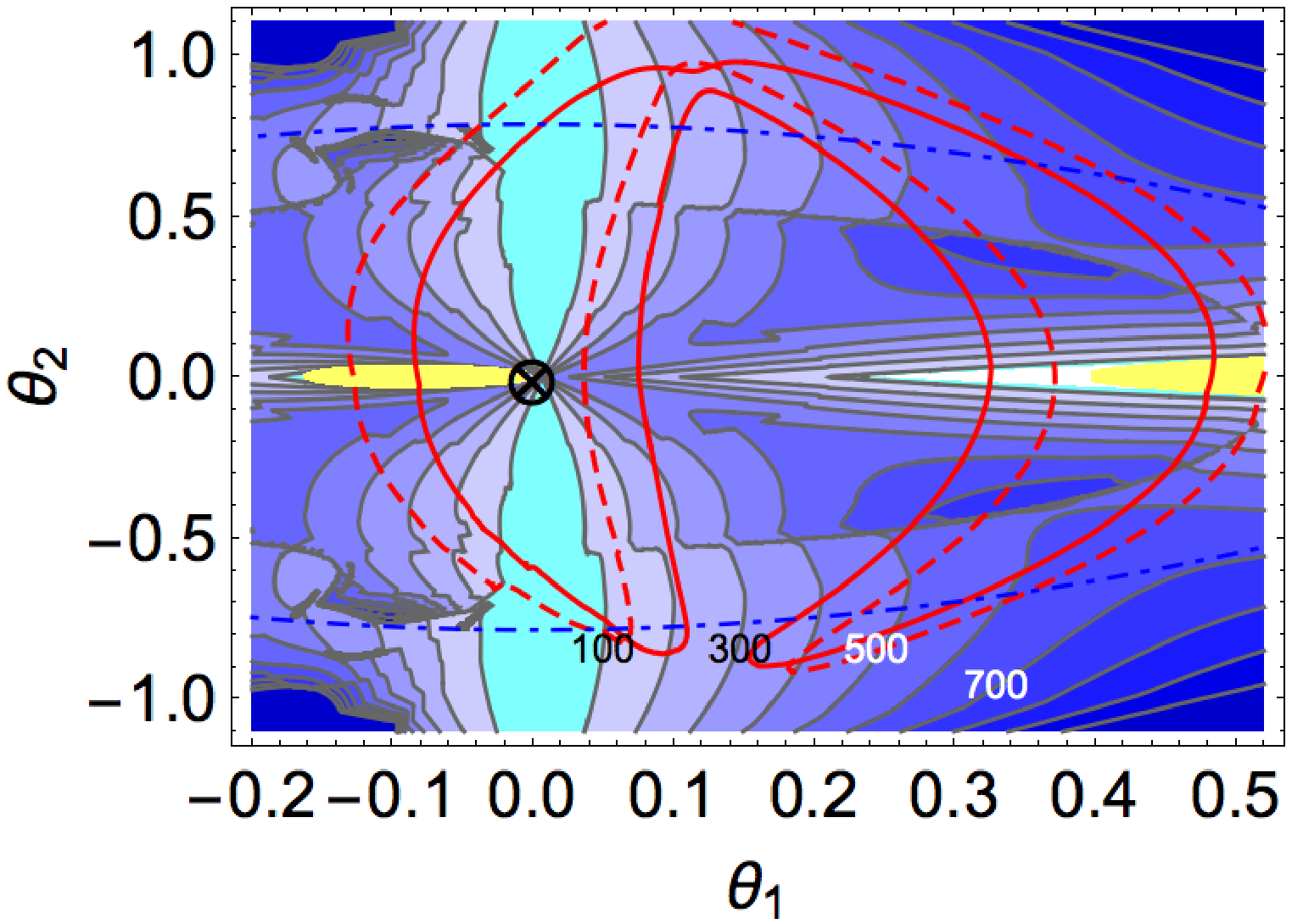,angle=0,width=7.8cm}}
{\hspace*{.2cm}\epsfig{figure=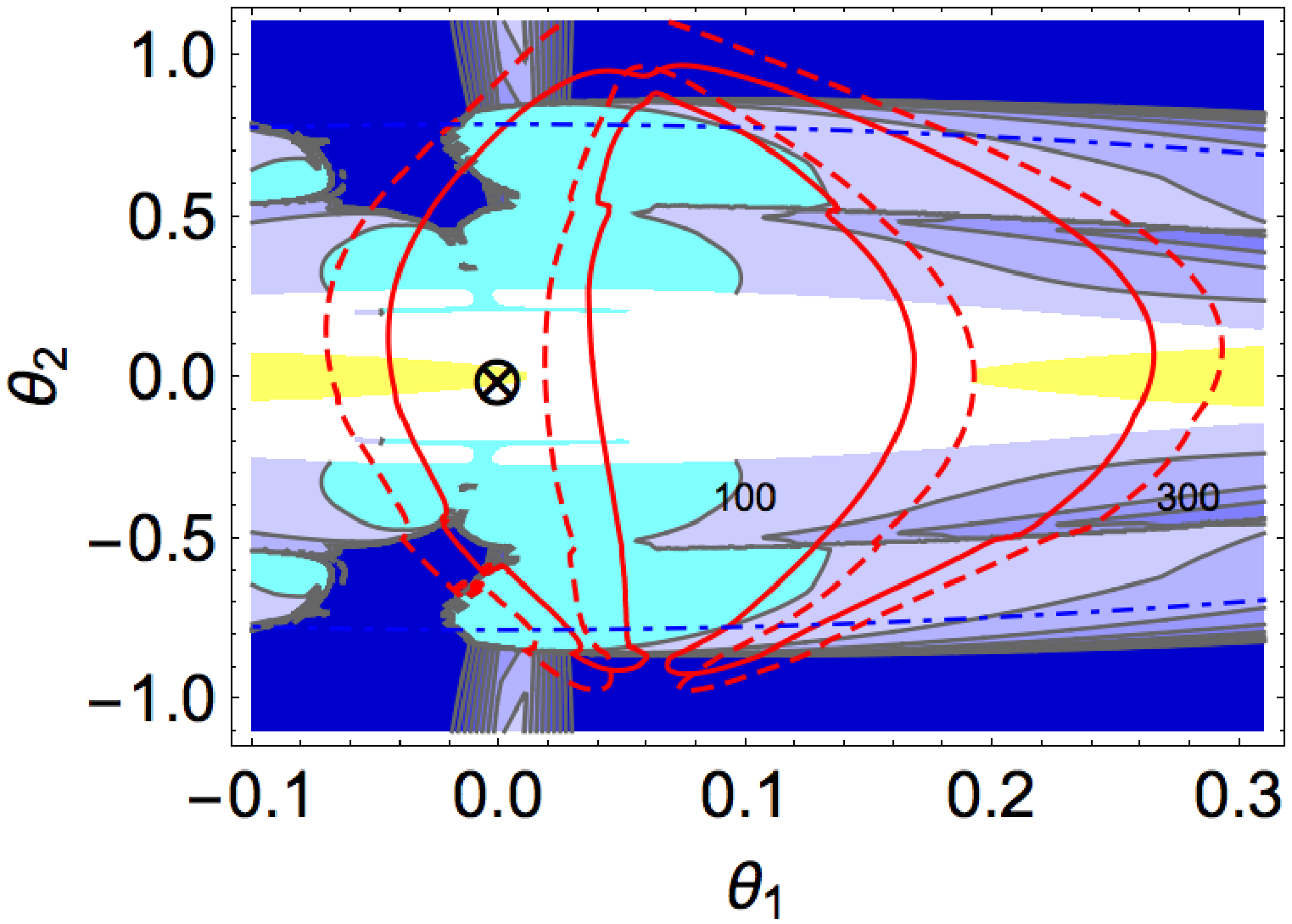,angle=0,width=7.8cm}}
}
\vspace{0.3cm}
\centerline{
{\hspace*{0cm}\epsfig{figure=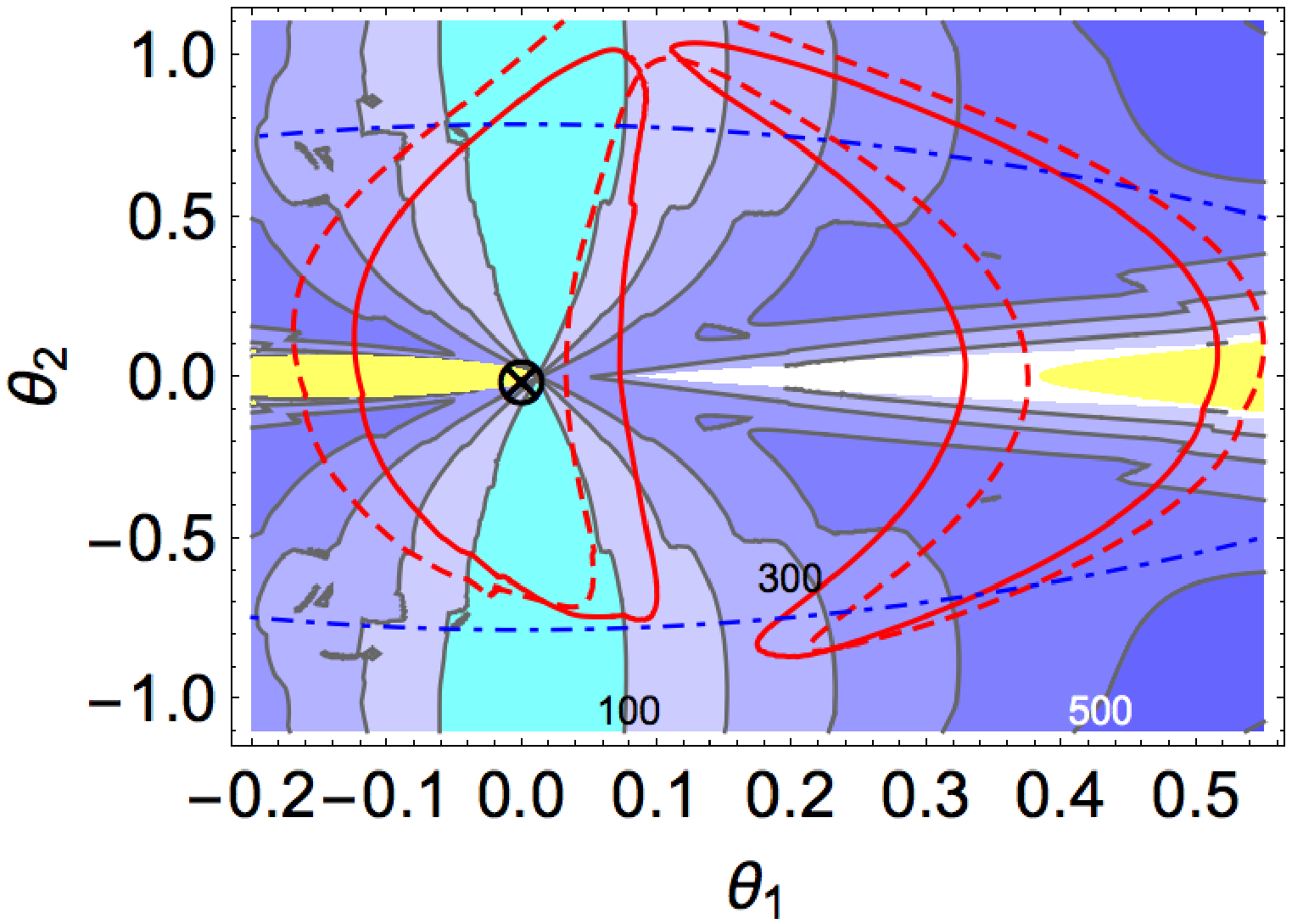,angle=0,width=7.8cm}}
{\hspace*{.2cm}\epsfig{figure=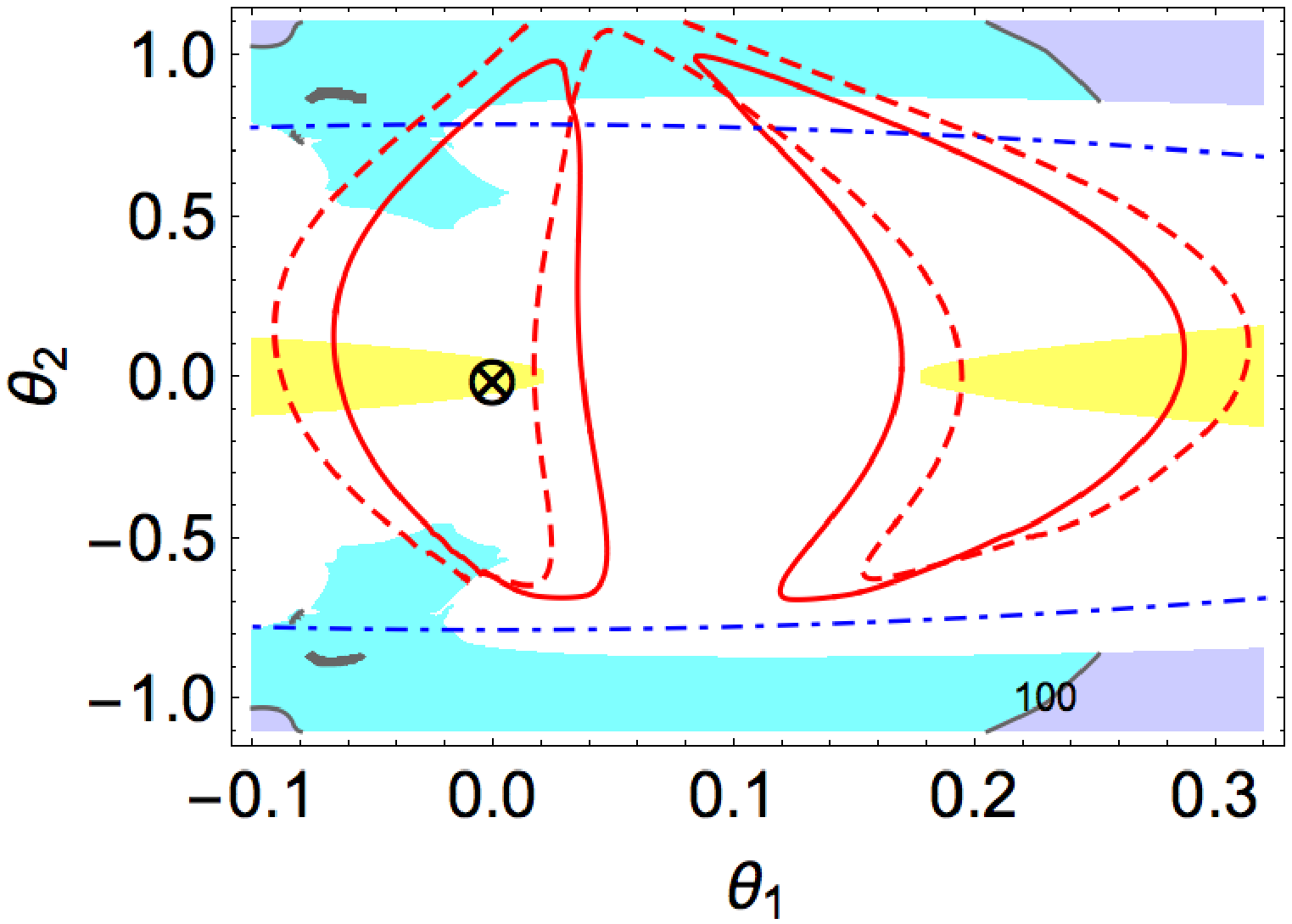,angle=0,width=7.8cm}}
}
\caption{
Upper bound on the higgsino mass parameter, $|\mu|_{\rm max}$, for $\tan\beta=5$ (left)
and $\tan\beta=10$ (right) in the NMSSM with $m_h=126$~GeV, $m_H=800$~GeV, and $m_s$
taken to maximize $\mu$ under the LEP constraint.
We have taken $m_0=120$~GeV ($115$~GeV) in the upper (lower) panel taking into account
that $m_0$ increases and the stop contribution to the $ggh$ coupling decreases
as the stop mass increases.
We plot the constant contours of $|\mu|_{\rm max}=100,\,200,\,300,\,400,\,\cdots$~GeV
on the $(\theta_1,\theta_2)$ plane.
The solid and dashed red circle are the region at preferred the $95$\% CL by the ATLAS and
CMS results, respectively.
The $\hat h$ fraction in $h$ is larger than $0.5$ in the region between two dot-dashed
blue curves.
The coupling $\lambda$ is above the perturbative bound in the unshaded region, while
$|\mu|_{\rm max}<100$~GeV in the cyan-shaded region.
The yellow-shaded region is excluded since there is no solution satisfying the relation
(\ref{NMSSM-relations-m0}).
}
\label{fig:result}
\end{minipage}
\end{center}
\end{figure}

Let us continue to examine the maximum value of $\mu$ in the region of $(\theta_1,\theta_2)$
compatible with the current LHC data.
This is done by varying $m_s$ for the given values of $m_0$, $m_h$, $m_H$ and $\tan\beta$.
As was done above, the relation (\ref{NMSSM-relations-m0}) is used to fix $\theta_3$ at each
point, and then the relations (\ref{NMSSM-relations-lam2}) and (\ref{NMSSM-relations-lamMu})
are combined to examine which value of $m_s$ maximizes $\mu$ under the LEP constraints
on $s$ if $m_s<114$~GeV.
Before going into the analysis, we present an approximated expression for the upper
bound on $\mu$:
\bea
|\mu|_{\rm max} \sim
350 \,{\rm GeV}\times\left(
\frac{2|\theta_1\theta_2|}{\theta^2_1+\theta^2_2}
\left(\frac{m_H}{800{\rm GeV}}\right)^{2}
\left(\frac{\tan\beta}{10}\right)^{-1}
+ 0.1 |\theta_2| \right),
\eea
which is obtained from (\ref{mu-upper-bound}) for $\theta_2\neq 0$.
Here $2|\theta_1\theta_2|/(\theta^2_1+\theta^2_2)\leq 1$, and it takes the maximum
value at $\theta_1=\pm \theta_2$.
The above expression is in the good agreement with the value evaluated from
(\ref{NMSSM-relations-m0}), (\ref{NMSSM-relations-lam2}) and (\ref{NMSSM-relations-lamMu})
in the region of $(\theta_1,\theta_2)$ where $\lambda$ is below the perturbative bound
and $h$ is SM-like, with $O^2_{h\hat h}>0.5$ and the signal rates compatible with
the current LHC data.
One can see that the upper bound on $\mu$ becomes stringent at large $\tan\beta$
and small $m_H$.
We find that the higgsinos have masses around or below $300$\,GeV
in the $95$\% preferred region by the LHC data on the Higgs signal rates, as long as
the heavy doublet Higgs boson has a mass, $m_H\lesssim 250\sqrt{\tan\beta}$~GeV,
and the singlet-like Higgs boson is consistent with the LEP constraints.
Also important is the dependence of $|\mu|_{\rm max}$ on $m_0$.
For fixed $m_\alpha$, the viable region with $\lambda<1$ is pushed far away from
the origin on the $(\theta_1,\theta_2)$ plane if one takes small $m_0$ or large $\tan\beta$,
because then the observed Higgs boson mass $m_h\simeq 126$~GeV requires large
$\lambda$ or large singlet-doublet mixing.
The Higgs coupling to photons receives a contribution from the charged-higgsino loops
according to (\ref{delta-Cr}), which can be sizable for $\mu$ below a few hundred
GeV in the region where the singlet-doublet mixing plays an important role in achieving
$m_h\simeq 126$~GeV.

Our results are summarized in Fig.~\ref{fig:result}.
The upper bound on the higgsino mass parameter, $|\mu|_{\rm max}$, is displayed
on the $(\theta_1,\theta_2)$ plane for the NMSSM with $m_h=126$~GeV and $m_H=800$~GeV,
taking $\tan\beta=5$ and $10$ in the left and right plot, respectively.
Here we have taken $m_0=120$~GeV ($115$~GeV) in the upper (lower) panel taking
into account two cases with stop mass above and below $600$~GeV as in (\ref{cases-stop-mass}).
Note that $m_0$ increases with the stop mass.
The constant contours of $|\mu|_{\rm max}=100,\,200,\,300,\,400,\cdots$~GeV
are shown in the gray lines.
The solid and dashed red circles correspond to the $95$\% preferred regions by
the ATLAS and CMS results, respectively.
In the yellow-shaded region there is no real value of $\theta_3$ satisfying the relation
(\ref{NMSSM-relations-m0}).
The value of $|\mu|_{\rm max}$ is less than $100$~GeV in the cyan-shaded region,
while $\lambda$ is larger than unity in the unshaded region.
In the region between two dot-dashed blue curves, the $\hat h$ fraction in $h$ is larger
than $0.5$, and $|\mu|_{\rm max}$ decreases if one takes larger $\tan\beta$, because it
is roughly proportional to $m^2_H/\tan\beta$.
We also see that smaller $m_0$ or larger $\tan\beta$ pushes the region with
$\lambda<1$ further away from the origin, thereby requiring smaller higgsino
mass in the viable region where $h$ is SM-like and has properties compatible
with the LHC results.
As explained above, this is because $m_h$ is the sum of $m_0$ and the additional
NMSSM contributions.
Finally it is worth noting that, if one allows $\lambda>1$ at the weak scale, the blue
and cyan area simply extend to the unshaded region.
There is no change in our conclusion that the higgsinos have masses around or below
$300$\,GeV for $m_H\lesssim 250\sqrt{\tan\beta}$~GeV in the region compatible with
the LHC and LEP constraints.

\subsection{CP-odd Higgs bosons}

Let us shortly discuss the CP-odd Higgs sector.
The lightest CP-odd neutral Higgs boson $A$ interacts with SM particles through
the doublet Higgs component, implying that there arise $Abb$, $ZhA$ and $ZsA$
coupling, but no $AWW$ and $AZZ$ coupling at the tree-level \cite{review-NMSSM}.
The Higgs boson $A$ obtains a mass according to
\bea
m^2_A = \frac{|b|}{\sin2\beta} + \frac{1}{2} m^{\prime 2}_{\hat s}
- \sqrt{
\left( \frac{|b|}{\sin2\beta} - \frac{1}{2} m^{\prime 2}_{\hat s} \right)^2
 + \lambda^2 v^2 A^2_\lambda
},
\eea
where $m^\prime_{\hat s}$ has a value different from $m_{\hat s}$ appearing
in the CP-even Higgs mass matrix (\ref{mass-squared}) because the singlet scalar
receives explicit U$(1)_S$ breaking mass contributions from the superpotential
$f(S)$ and the associated soft SUSY breaking terms.

The case of our interest is that $A$ is singlet-like, and the doublet-like
CP-odd Higgs boson is much heavier than $A$.
Then the $hAA$ coupling is approximately given by $\lambda^2 v$, and the $Abb$
couplings is estimated as
\bea
y_{Abb} \simeq y_b \tan\beta \sin\phi,
\eea
with $y_b$ being the bottom quark Yukawa coupling.
The mixing angle $\phi$ between CP-odd Higgs bosons is smaller than about $m^2_s/m^2_H$,
and the $ZhA$ and $ZsA$ couplings vanish in the decoupling regime where one combination
of $H_u$ and $H_d$ is much heavier than the weak scale.
There are LEP constraints on the processes, $e^+e^-\to ZA \to Zb\bar b$, and
$e^+e^- \to Z^* \to sA$ or $hA$, depending on the mass of $A$.
Using the properties discussed above, one finds that these constraints can be avoided
without difficulty when $A$ is singlet-like.
On the other hand, the Higgs signal rate at the LHC is modified by the process,
$h\to A A^* \to 4b$, if kinematically open.
The branching fraction of this decay mode is however smaller than the decay via
$h\to ZZ^* \to 4b$ for $y_{Abb}\ll 1$ and $\lambda<1$.

\subsection{Neutralino sector}

The NMSSM neutralino sector includes the singlino, which modifies the property of
the lightest neutralino crucially depending on the supersymmetric singlino mass.
The singlet superpotential is written as
\bea
f(S) = \xi S + m S^2 + \kappa S^3,
\eea
neglecting terms suppressed by the cut-off scale of the theory.
Here the tadpole and mass terms should be around or below TeV to achieve EWSB
without severe fine-tuning.
These terms are suppressed if one imposes a discrete symmetry such
as $Z_3$, but with small explicit breaking so as to avoid the domain-wall problem
\cite{review-NMSSM}.
Another interesting and natural way is to incorporate the Peccei-Quinn symmetry solving
the strong CP problem via the invisible axion, so that $S$ obtains small tadpole and
mass terms only after the Peccei-Quinn symmetry is spontaneously broken
\cite{Jeong:2011jk}.

The lightest neutralino $\chi$ interacts with the SM particles, and there are
various experimental constraints on its couplings, in particular, on those to
the SM-like Higgs boson and the $Z$-boson:
\bea
{\cal L}_{\rm int} = \frac{1}{2} \left( y_\chi h \bar\psi\psi + {\rm h.c.} \right)
+ c_\chi \frac{m_Z}{v} \bar\psi \gamma^\mu \gamma^5 \psi Z_\mu
+ \cdots,
\eea
where $\psi^T=(\chi,\bar\chi)$ is the four-component spinor, and
the couplings are determined by the neutralino mixing parameters
\bea
y_\chi &=& \Big(
g^\prime N_1(N_4\sin\beta-N_3\cos\beta)
- g N_2 (N_4\sin\beta -N_3 \cos\beta) \Big)\cos\theta_1\cos\theta_2
\nonumber \\
&&
-\,\sqrt2 \lambda N_5(N_4\cos\beta+N_3\sin\beta)\cos\theta_1\cos\theta_2,
\nonumber \\
c_\chi &=& N^2_3 - N^2_4,
\eea
in the presence of scalar mixing.
Here the lightest neutralino $\chi$ is composed by
\bea
\chi = N_1 \tilde B + N_2 \tilde W^3 + N_3 \tilde H^0_d + N_4 \tilde H^0_u
+ N_5 \tilde S,
\eea
with $g$ and $g^\prime$ being the SU$(2)$ and U$(1)_Y$ gauge couplings, respectively.
The mixing parameters are fixed by diagonalizing the mass matrix
\bea
\begin{pmatrix}
M_{\tilde B} & 0 & -\frac{g^\prime v}{\sqrt2} \cos\beta & \frac{g^\prime v}{\sqrt2} \sin\beta & 0 \\
0 & M_{\tilde W} & \frac{g v}{\sqrt2} \cos\beta & -\frac{g v}{\sqrt2}\sin\beta & 0 \\
-\frac{g^\prime v}{\sqrt2}\cos\beta & \frac{g v}{\sqrt2}\cos\beta & 0 & -\mu & -\lambda v \sin\beta \\
\frac{g^\prime v}{\sqrt2} \sin\beta & -\frac{g v}{\sqrt2}\sin\beta & -\mu & 0 & -\lambda v \cos\beta \\
0 & 0 & -\lambda v \sin\beta & -\lambda v\cos\beta & \langle \partial^2_S f \rangle
\end{pmatrix},
\eea
which is given in the basis, $(\tilde B,\tilde W,\tilde H^0_d,\tilde H^0_u, \tilde S)$,
with $M_{\lambda}$ being the mass of the indicated gaugino.

Let us briefly discuss the constraints on the neutralino sector.
Since the NMSSM with a light singlet scalar requires relatively light higgsinos, we pay
our attention to the case where the lightest neutralino has a sizable higgsino component.
The $h\chi\chi$ and $Z\chi\chi$ coupling are constrained by the LUX and XENON results
on direct dark matter searches \cite{LUX,Cohen:2010gj}
\bea
\sigma_{\rm SI} &\simeq& 0.9\times 10^{-44}\,{\rm cm}^2
\left(\frac{y_\chi}{0.1}\right)^2
\left(\frac{m_h}{126{\rm GeV}}\right)^{-4}
\,\lesssim\, 0.8\times 10^{-45}\,{\rm cm}^2 \left(\frac{\Omega_\chi h^2}{0.11}\right)^{-1},
\\
\sigma_{\rm SD} &\simeq& 0.8\times 10^{-40}\,{\rm cm}^2
\left(\frac{c_\chi}{0.1}\right)^2
\,\lesssim\, 0.35\times 10^{-39}\,{\rm cm}^2 \left(\frac{\Omega_\chi h^2}{0.11}\right)^{-1},
\eea
for $\Omega_\chi h^2$ being the relic energy density of $\chi$.
Here the upper limit on the spin-independent neutralino-nucleon cross section
is for $m_\chi=33$~GeV \cite{LUX},
where it reaches the minimum, while the upper
limit on the spin-dependent one is for $m_\chi=45$~GeV \cite{Aprile:2012nq}.
If $\chi$ constitutes the main component of dark matter, the above requires
both the $h\chi\chi$ and $Z\chi\chi$ coupling to be smaller than about $0.1$ unless
$\chi$ is lighter than $10$~GeV.
For the NMSSM with relatively light higgsinos, such small couplings are obtained
if $\chi$ is almost higgsino-like, or if the singlino or bino is
much lighter than the higgsino.
The experimental constraints from the direct dark matter search are relaxed
if $\chi$ composes a portion of the dark matter,
for which one may consider the gravitino, axino, and/or axion as the main component
of dark matter.
The $h\chi\chi$ coupling is further constrained by the LHC bound on the Higgs invisible
decay if $2m_\chi<m_h$ \cite{Giardino:2013bma}.
In addition, the LEP experiment puts a constraint on the neutralino production if
the sum of the lightest and the second lightest neutralino masses is below
$209$~GeV \cite{OPAL}.

We close this subsection by mentioning the relic abundance of the lightest neutralino
in the NMSSM with R-parity conservation.
If $\chi$ has a large higgsino component, the t-channel chargino-mediated
process $\chi\chi\to W^+W^-$ occurs with a large annihilation cross section
for $m_\chi>m_W$ \cite{Jungman:1995df}, and thus the dark matter of the Universe
cannot be explained by the neutralino thermal relic alone.
The process $\chi\chi\to hh$ or $ss$ can be also important if $m_\chi>m_h$.
To get a sufficient relic density, one may consider sizable mixing with bino
or singlino.
Another way is to consider non-thermal production, or other dark matter candidates
such as the gravitino, axino, and/or the axion.
On the other hand, for the case where $\chi$ has a mass below $m_W$ but above $m_h/2$,
the thermal relic abundance of $\chi$ is too large if the s-channel $Z$-boson exchange
dominates the neutralino annihilation.
This can be avoided by Higgs resonant annihilation.
One may instead rely on non-thermal production, or late-time entropy production
diluting the neutralino abundance.

\section{Conclusions}\label{sec:conclusion}

The SM-like Higgs boson discovered at the LHC places important constraints on
the supersymmetric extensions of the SM.
Extended to include a gauge singlet, the Higgs sector can naturally
explain the observed Higgs boson mass within TeV scale SUSY.
In this paper we have focused on the NMSSM scenario where the singlet scalar is below
the weak scale so that singlet-doublet mixing enhances the mass of the SM-like
Higgs boson, and examined the phenomenological consequences of scalar mixing.
The current experimental data allows sizable scalar mixing in the region around
$(\theta_1,\theta_2)=(0,0)$ and $(2/\tan\beta,0)$.
The two regions are distinguished by the sign of the Higgs coupling to down-type
fermions.

The higgsino mass parameter and the singlet coupling to Higgs bilinear have a
crucial dependence on the Higgs boson masses and mixing angles.
Using the relations among them we found that the scalar mixing compatible with
the LHC results on the SM-like Higgs boson leads to relatively light higgsinos,
around or below a few hundred GeV, as long as the heavy doublet Higgs boson has
a mass smaller than about $250\sqrt{\tan\beta}$~GeV, and the singlet-like Higgs
boson is consistent with the LEP constraints.
Also important is that the charged-higgsino loops combined with singlet-doublet
mixing give a contribution to the Higgs coupling to photon,
which has either sign and can be sizable when the higgsinos are light.
The future run of the LHC and future linear collider experiments will clarify
the viable range of mixing with higher accuracy, and could detect
the singlet-like Higgs boson while probing the structure of the Higgs sector.

\begin{acknowledgements}

KSJ is supported by IBS under the project code, IBS-R018-D1. YS is supported by Grant-in-Aid for JSPS Fellows under the program number
26$\cdot$3171, and Tohoku University Institute for International Advanced Research
and Education.
MY is supported by Grants-in-Aid for Scientific Research
from the Ministry of Education, Science, Sports, and Culture (MEXT), Japan,
No.~23104008 and No.~23540283.

\end{acknowledgements}

\end{document}